\documentclass[useAMS,usenatbib]{mnras}
\usepackage{graphicx}
\usepackage[english]{babel}
\usepackage{amsmath}
\usepackage{amssymb}
\usepackage{natbib}
\usepackage{wasysym}
\usepackage{newtxtext,newtxmath}
\usepackage[T1]{fontenc}
\usepackage{url}
\usepackage{xcolor}
\usepackage{dblfloatfix}

\title[To $B$ or not to $B$]{The effect of magnetic fields on properties of the circumgalactic medium}
\author[F. van de Voort et al.]{Freeke~van~de~Voort,$^{1,2}$\thanks{E-mail: freeke@astro.cf.ac.uk} Rebekka Bieri,$^2$ R\"udiger Pakmor,$^2$ Facundo~A.~G\'omez,$^{3,4}$
\newauthor
Robert~J.~J.~Grand$^2$ and Federico Marinacci$^5$\\
$^1$School of Physics \& Astronomy, Cardiff University, Queens Buildings, The Parade, Cardiff CF24 3AA, UK \\
$^2$Max Planck Institute for Astrophysics, Karl-Schwarzschild-Stra{\ss}e 1, 85748, Garching, Germany \\
$^3$Instituto de Investigaci\'on Multidisciplinar en Ciencia y Tecnolog\'ia, Universidad de La Serena, Ra\'ul Bitr\'an 1305, La Serena, Chile \\
$^4$Departamento de F\'isica y Astronom\'ia, Universidad de La Serena, Av. Juan Cisternas 1200 Norte, La Serena, Chile \\
$^5$Department of Physics and Astronomy, University of Bologna, via Gobetti 93/2, 40129 Bologna, Italy
}

\begin{document}

\date{Accepted 2020 December 18. Received 2020 December 17; in original form 2020 August 17}

\pagerange{\pageref{firstpage}--\pageref{lastpage}} \pubyear{2020}

\maketitle

\label{firstpage}

\begin{abstract}

\noindent
We study the effect of magnetic fields on a simulated galaxy and its surrounding gaseous halo, or circumgalactic medium (CGM), within cosmological `zoom-in' simulations of a Milky Way-mass galaxy as part of the `Simulating the Universe with Refined Galaxy Environments' (SURGE) project. We use three different galaxy formation models, each with and without magnetic fields, and include additional spatial refinement in the CGM to improve its resolution. The central galaxy's star formation rate and stellar mass are not strongly affected by the presence of magnetic fields, but the galaxy is more disc-dominated and its central black hole is more massive when $B>0$. The physical properties of the CGM change significantly. With magnetic fields, the circumgalactic gas flows are slower, the atomic hydrogen-dominated extended discs around the galaxy are more massive and the densities in the inner CGM are therefore higher, the temperatures in the outer CGM are higher, and the pressure in the halo is higher and smoother. The total gas fraction and metal mass fraction in the halo are also higher when magnetic fields are included, because less gas escapes the halo. Additionally, we find that the CGM properties depend on azimuthal angle and that magnetic fields reduce the scatter in radial velocity, whilst enhancing the scatter in metallicity at fixed azimuthal angle. The metals are thus less well-mixed throughout the halo, resulting in more metal-poor halo gas. These results together show that magnetic fields in the CGM change the flow of gas in galaxy haloes, making it more difficult for metal-rich outflows to mix with the metal-poor CGM and to escape the halo, and therefore should be included in simulations of galaxy formation. 
\end{abstract}

\begin{keywords}
methods: numerical -- magneto-hydrodynamics -- galaxies: magnetic fields -- galaxies: formation -- galaxies: haloes -- intergalactic medium 
\end{keywords}

\section{Introduction}

The gaseous component of the dark matter-dominated haloes around galaxies is also referred to as the circumgalactic medium (CGM). It plays an important role in the formation of galaxies. Galaxies grow by accreting gas from their surrounding haloes. Gas stripped from galaxies or ejected from them by feedback is returned to the CGM, thereby heating it and enriching it with heavy elements. The physical properties of the CGM thus determine in part how galaxies grow and evolve and they yield information about feedback processes and their resulting outflows \citep[e.g.][]{Voort2012, Suresh2015}. Understanding its properties, and how they depend on different physical processes, is therefore vital for theories of galaxy formation as well as for providing reliable predictions for and interpretations of observations of the CGM \citep[e.g.][]{Putman2012, Tumlinson2017}. 

The gas around galaxies is known to have a range of temperatures and metallicities \citep{Werk2014, Prochaska2017, Lehner2019}. These properties correlate with one another. For example, cosmological simulations find that inflowing gas is cooler and denser than outflowing gas, on average \citep[e.g.][]{Voort2012, Ford2014}. Simulated CGM properties also depend on halo mass, on redshift, and on the implementation of feedback from stars and supermassive black holes \citep{Voort2012, Suresh2015}. The latter makes it a challenge for simulations to reproduce observations of the cool, intermediate temperature, and hot CGM \citep{Hummels2013, Gutcke2017, Jakobs2018,  Ji2020}. The consequences of additional physical processes on the CGM are therefore being explored. Examples of such processes are cosmic ray feedback and thermal conduction, which both depend on the magnetic field strength and orientation \citep[e.g.][]{Barnes2019, Buck2020, Ji2020}. Here, we focus on the relatively unexplored effect of the magnetic fields themselves and study how they may change the physical properties of circumgalactic gas at low redshift. 

Magnetic fields are ubiquitous in the Universe \citep[e.g.][and references therein]{Beck2013} and are believed to be important for a variety of astrophysical processes, such as for star formation and for launching jets from protostars or black holes \citep[e.g.][and references therein]{Tchekhovskoy2015, Hennebelle2019, Girichidis2020}. Star-forming disc galaxies are also observed to have large-scale magnetic fields, not just in their interstellar medium \citep[ISM, e.g.][]{PlanckXXXV2016, Han2017}, but also in regions around the disc \citep[e.g.][]{Haverkorn2012, Han2017}. The interpretation of observations of extraplanar magnetic fields is not straightforward, but their orientation may indicate a correlation with the flow of gas in the inner halo \citep{Ferriere2014}. 

State-of-the-art simulations of galaxy formation do not always include magnetic fields \citep[e.g.][]{Dubois2014, Schaye2015, Tremmel2017, Hopkins2018, Dave2019}. However, recently there has been a push to include them in cosmological simulations, firstly as a diagnostic tool in comparison with observations and secondly because they are required for additional physical processes, such as cosmic ray transport and thermal conduction \citep[e.g.][]{Dubois2008, Pakmor2017, Rieder2017, Marinacci2018, MartinAlvarez2018, Hopkins2020}. In these simulations, the magnetic field is amplified strongly by a turbulent dynamo in the ISM and pushed into the CGM by galactic outflows. Some studies find that they do not have an important dynamical effect on galaxies \citep{Pakmor2017, Su2017, Hopkins2020}, whereas others find that they have can noticeably change certain galaxy properties, such as galaxy sizes \citep{Pillepich2018, MartinAlvarez2020}. Although different numerical methods and magnetic field implementations may lead to different results, these seemingly conflicting conclusions may be explained simply by the authors focusing on different aspects of galaxy formation or on different redshifts. 

Studies agree that, on average, the thermal pressure dominates over magnetic pressure in the CGM. However, even though not dominant, magnetic fields may well be dynamically important \citep{Ji2018}. Moreover, even though most of the halo volume is filled with hot, diffuse gas with high thermal pressure, some of the CGM is much cooler than the virial temperature and can be dominated by magnetic pressure \citep{Nelson2020}. A more in-depth investigation into how magnetic fields interact with the multiphase CGM, especially at low redshift, is clearly warranted. We therefore study the density, temperature, pressure, metallicity, and radial velocity of the gas around a Milky Way-mass galaxy and find substantial differences depending on the presence or absence of magnetic fields, both in regions where the magnetic pressure is dominant and where it is sub-dominant. 

In this work, we use cosmological `zoom-in' simulations of a single Milky Way-mass halo with three different variations of the subgrid physics and with enhanced resolution in the CGM as part of the `Simulating the Universe with Refined Galaxy Environments' (SURGE) project. We study the impact of the magnetic field on properties of the CGM and the central galaxy. We show that the gas flows in the halo are substantially altered by the inclusion of magnetic fields, even on large scales where the magnetic pressure is subdominant.
The simulation method is briefly described in Section~\ref{sec:sim}. In Section~\ref{sec:results} we present our results on magnetohydrodynamic properties of the CGM. We summarize and discuss our findings in Section~\ref{sec:concl}. Appendix~\ref{sec:test} shows results for standard resolution simulations from which we conclude that our results are independent of resolution. Appendix~\ref{sec:haloes} compares results from standard resolution simulations with different initial conditions to study halo-to-halo variation. Finally, the full distribution of CGM properties are shown in Appendix~\ref{sec:pdf}. All length scales are given in proper coordinates.

\section{Method} \label{sec:sim}

\begin{figure*}
\center
\includegraphics[scale=0.7]{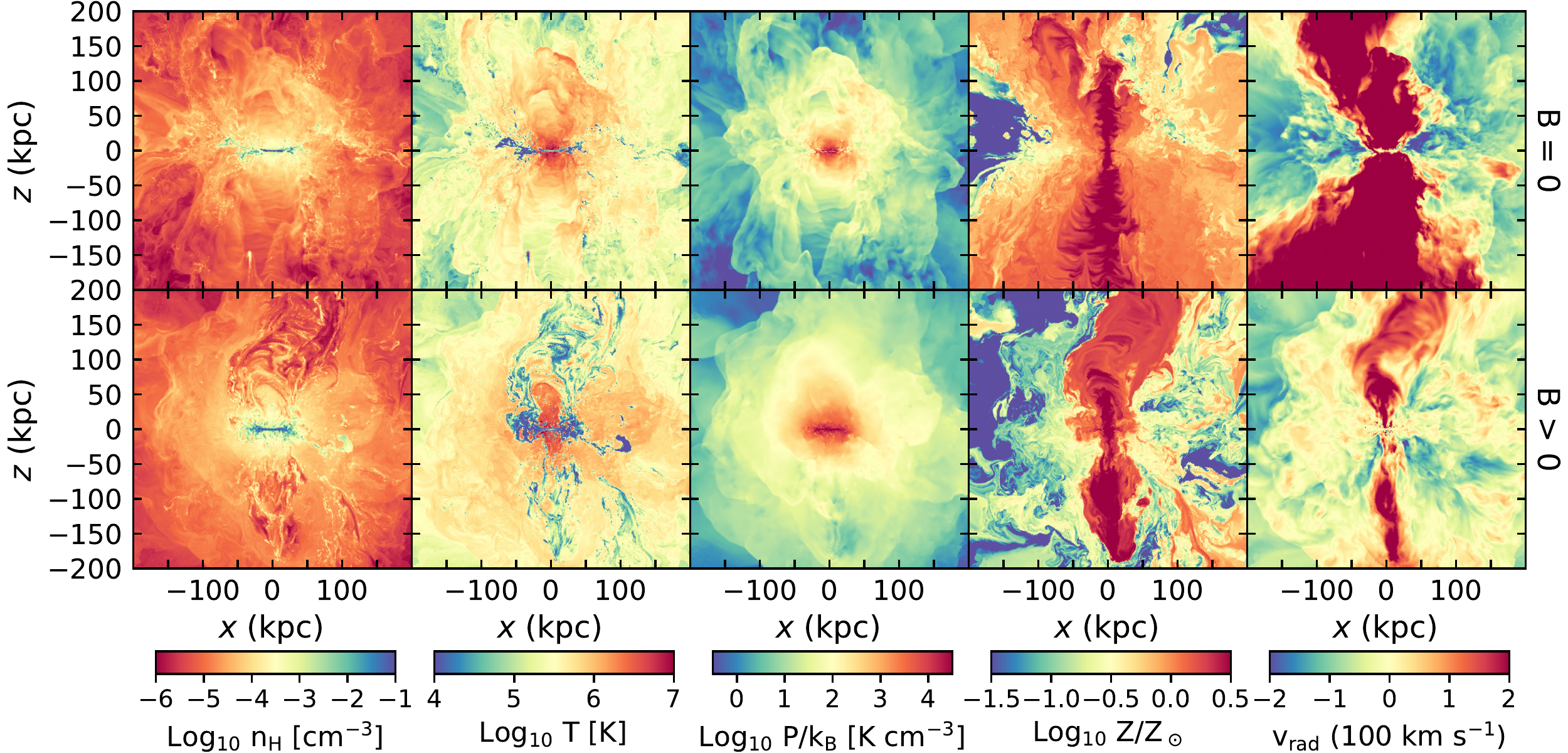}
\caption {\label{fig:imgBnoB} $400\times400$~kpc$^2$ images of the gas in and around an edge-on Milky Way-mass galaxy at $z=0$ in a simulation without magnetic fields (top panels) and with magnetic fields (bottom panels). From left to right the panels show the gas density, its temperature, the total pressure (thermal pressure plus magnetic pressure), the metallicity, and the radial velocity of the gas. The top (bottom) panels show the `Auriga noAGN' simulation without (with) magnetic fields. All panels show an infinitesimally thin slice through the centre of the galaxy. In general, the density exhibits smaller scale fluctuation without $B$. The cool gas in the CGM ($T<10^5$~K) has more coherent, filamentary structure with $B$. The pressure in the halo is also smoother with $B$. The metallicity is highest along the polar direction in both cases, but the other regions of the CGM are much more enriched without $B$, indicating that the gas mixes more efficiently. The radial velocity is strongly time variable, but generally shows more confined outflows when $B>0$.}
\end{figure*} 

This work is an extension of the Auriga project\footnote{\url{https://wwwmpa.mpa-garching.mpg.de/auriga/}} \citep{Grand2017}, which consists of a large number of zoom-in magnetohydrodynamical, cosmological simulations of relatively isolated Milky Way-mass galaxies and their environments. We resimulate one of the Auriga galaxies (halo~6) at standard mass resolution with additional spatial resolution in the CGM.\footnote{Two other Milky Way-mass haloes (halo~12 and halo~L8) are shown and discussed in Appendix~\ref{sec:haloes}.} Halo~6 was chosen, because the central galaxy has properties that are reasonably similar to the Milky Way. Additionally, it has a relatively compact zoom-in region, which makes it one of the most efficient in terms of computational time. Our simulations are part of the SURGE project (see also \citealt{Voort2019}). 

In order to check that our results are robust to changes in the employed feedback model, we run three variations:
\begin{enumerate}
\item the Auriga model without active galactic nucleus (AGN) feedback, `Auriga noAGN';
\item the Auriga model, `Auriga';
\item the IllustrisTNG model, `TNG'. 
\end{enumerate}
The latter two models both include AGN feedback, but with different implementations (see below).

All of the simulations were run with the latest version of the quasi-Lagrangian moving mesh code \textsc{arepo} \citep{Springel2010, Pakmor2016, Weinberger2020} and assume a $\Lambda$CDM cosmology with parameters taken from \citet{PlanckXVI2014}: $\Omega_\mathrm{m}= 1-\Omega_\Lambda = 0.307$, $\Omega_\mathrm{b}= 0.048$, $h = 0.6777$, $\sigma_8 = 0.8288$, and $n = 0.9611$. \textsc{Arepo} uses a second order finite volume scheme on an unstructured Voronoi mesh for the gas. Dark matter, stars, and black holes are modelled as collisionless particles. All of our models include ideal magnetohydrodynamics (MHD) \citep{Pakmor2013, Pakmor2014} and use the Powell scheme for divergence control \citep{Powell1999}. The relative divergence error is typically a few per cent \citep{Pakmor2020}. Primordial and metal-line cooling with self-shielding corrections \citep{Vogelsberger2013, Rahmati2013} and a time-dependent ultra-violet (UV) background \citep{Faucher2009} are included as well. The fraction of atomic hydrogen (H~\textsc{i}) is calculated on-the-fly based on the UV background radiation, the local radiation field from the AGN if included, and the self-shielding approximation from \citet{Rahmati2013}.

Stellar mass-loss and metal return is included for core-collapse and Type Ia supernovae and for asymptotic giant branch stars based on tabulated mass and metal yields \citep[and references therein]{Grand2017, Pillepich2018}. We chose to inject mass and metals ejected by a star particle only into its host cell, rather than into its 64 neighbours as done in the original Auriga and IllustrisTNG models, but this change is unlikely to affect any of our results (only the $2\sigma$ scatter in stellar abundances is slightly higher, as shown in \citealt{Voort2020}). 

The mass resolution of our cells is the same as in the standard Auriga simulations, i.e.\ the target cell mass for baryons is $5.4\times10^4$~M$_{\astrosun}$ and dark matter particle masses are $2.9\times10^5$~M$_{\astrosun}$.  At $z=0$, all cells within 1~Mpc of the central Milky Way-mass galaxy are within the `zoom-in' region and therefore are refined to this target mass. Additionally, following \citet{Voort2019}, we limit the cell volume to a maximum volume of (1~kpc)$^3$ for cells within $1.2R_\mathrm{vir}$ of any halo more massive than $10^{7.9}$~M$_\odot$ and located within the `zoom-in' region. In this work, we define the virial radius, $R_\mathrm{vir}$, as the radius within which the mean overdensity is 200 times the \emph{critical} density of the Universe at its redshift, where $R_\mathrm{vir}\approx210$~kpc at $z=0$. The total halo mass within this radius is $10^{12.0}$~M$_{\astrosun}$ at $z=0$.

The haloes are identified on-the-fly by running \textsc{subfind} \citep{Springel2001, Dolag2009} at each of the 128 output redshifts between $z=47$ and $z=0$. Sufficiently massive Friends-of-Friends (FoF) haloes are selected and `dyed' with a passive `refinement' scalar of value unity. Between two consecutive outputs, the refinement scalar is advected with the fluid. A cell is spatially refined (as described in \citealt{Springel2010}) if its refinement scalar is above 90 per cent of the injected value. Due to inflows and outflows, the spatial spherical refinement region becomes more deformed with time. The scalar is therefore reinitialized at the next output redshift. See \citet{Voort2019} for more details. For some of our results, we use \textsc{subfind} to identify bound substructures within FoF haloes in order to exclude satellite galaxies and the gas associated with them, but this does not impact our conclusions.

Star formation takes place stochastically in gas with densities above $n_\mathrm{H}^\star=0.11$~cm$^{-3}$ following \citet{Springel2003}. This model was calibrated to reproduce the Kennicutt-Schmidt law \citep{Kennicutt1998} in simulations without magnetic fields and was not recalibrated for our simulations, even though some of them include additional magnetic pressure. This is unimportant for most of the galaxy's evolution, because the magnetic pressure is subdominant \citep[see][]{Pakmor2017}. At low redshift, however, the magnetic pressure becomes more important and may change the density profile in the ISM somewhat. We will discuss this further in Section~\ref{sec:prop}. Generally, the entire gas cell is converted to a star particle, unless the total mass exceeds the target mass resolution by more than a factor of two, in which case only the target mass is converted to a star and the gas cell is retained with reduced mass. Because the multiphase structure of the ISM cannot be resolved at the resolution of our simulations, this gas is placed on an effective equation of state. Here, the galaxies' ISM is defined to be all of the star-forming gas (i.e.\ all gas with $n_\mathrm{H}>0.11$~cm$^{-3}$).

All models include stellar feedback and the `Auriga' and `TNG' models include both stellar and AGN feedback, which result in large-scale outflows (see \citealt{Grand2017}, \citealt{Weinberger2017}, and \citealt{Pillepich2018} for details). The stellar feedback models are quite similar in the different models. Briefly, `wind particles' are launched stochastically from the star-forming gas and are temporarily decoupled from hydrodynamic interactions, until they reach 5 per cent of $n_\mathrm{H}^\star$. The AGN feedback models differ substantially from each other. The feedback in both AGN models is most efficient when the central black hole has a low accretion rate. However, the implementations for this `radio mode' are quite different. In the case of `Auriga', energy from the AGN is added to the CGM as thermal energy, whereas in the case of `TNG', the energy is added directly around the supermassive black hole (in the ISM) as kinetic energy. In the `TNG' model, the AGN-driven outflows interact with the ISM, escape the disc along the path of least resistance, and are therefore more bipolar near the disc than those in the `Auriga' model.

For each of our three model variations we run two simulations, one with and one without magnetic fields. This is done by using an identical version of the code, but setting the seed magnetic field to zero instead of the fiducial $1.6\times10^{-10}$~physical~G, which is initialized as a uniform field at the start of the simulation, at $z=127$. The simulation self-consistently follows the same ideal MHD equations for the evolution of the gas regardless of whether or not the magnetic field was seeded. Note that we do not expect galaxy evolution to be affected by the exact value of the nonzero seed field in our model \citep{Marinacci2016}. Additionally, the choice of using a simple uniform seed field instead of more physically motivated seed fields is unlikely to matter at low redshift, as shown by \citet{Garaldi2020}. We ran the same six simulations with identical mass resolution, but without spatial refinement in the CGM, for our resolution tests shown in Appendix~\ref{sec:test}. We ran another four simulations without spatial refinement to check whether our results hold for other Milky Way-mass haloes besides halo~6, using the initial conditions of halo~12 \citep{Grand2017} and halo~L8 \citep{Grand2019}. These are shown and discussed in Appendix~\ref{sec:haloes}. All of these simulations have differently timed outflows, driven by starbursts or AGN, but each simulation shows the same qualitative results. We therefore believe that our results do not depend on such stochastic processes and likely hold in general for Milky Way-mass haloes.

\section{Results} \label{sec:results}


We found that the CGM properties are most strongly affected by the presence or absence of magnetic fields in the simulations without AGN feedback (`Auriga noAGN'; quantified in Sections~\ref{sec:prop} and~\ref{sec:prof}). We therefore use this model as our fiducial model, but show many of our results for the other models as well. Even though there are quantitative differences in the behaviour, the effect of the magnetic fields is significant and qualitatively the same in all models.

\subsection{A qualitative view}

Images of density, temperature, pressure, metallicity, and radial velocity in an infinitesimally thin slice through the CGM at $z=0$ are shown in Figure~\ref{fig:imgBnoB} for model `Auriga noAGN'. The galaxy was rotated such that its angular momentum lies along the $z$-axis and is therefore edge-on. The top panels show the gas properties in a simulation without magnetic fields and the bottom panels in one with $B$ fields. 

The density is higher near the disc in the presence of $B$ fields. There are more dense, cool clumps just above and below the disc and the cool gas in the plane of the disc between 20~and 50~kpc is more puffed up. This is the case for all of our models. At larger distances, the average density is similar, but the structure is quite different. The simulation with magnetic fields shows more coherent, filamentary structure everywhere in the halo, whereas the one without $B$ shows more fine-grained structure and appears more turbulent on small scales, possibly because there is more mixing, as discussed below. This is also the case for the other two models, although this structure difference is slightly weaker in the `Auriga' model, because its feedback is less directional than in `Auriga noAGN' or `TNG'. 

The volume is dominated by hot gas and its average temperature is slightly higher with magnetic fields. However, especially noticeable are the large filaments of cool, yet relatively low-density gas in the polar directions. This may be opposite to expectations given that this is the direction of the outflow, which is generally hot. Because this gas was (or is) part of a galactic outflow, it has a high metallicity, which allows it to cool efficiently. Feedback can disrupt this cool (but low-density) gas relatively easily, because it is located along the bipolar outflow direction. Our simulations with AGN feedback also show cool gas structures in the outflow direction, but they are much less prominently visible. These low-density filaments do not dominate the overall mass budget of cool gas (as shown quantitatively in Appendix~\ref{sec:pdf}). 

\begin{figure}
\center
\includegraphics[scale=0.8]{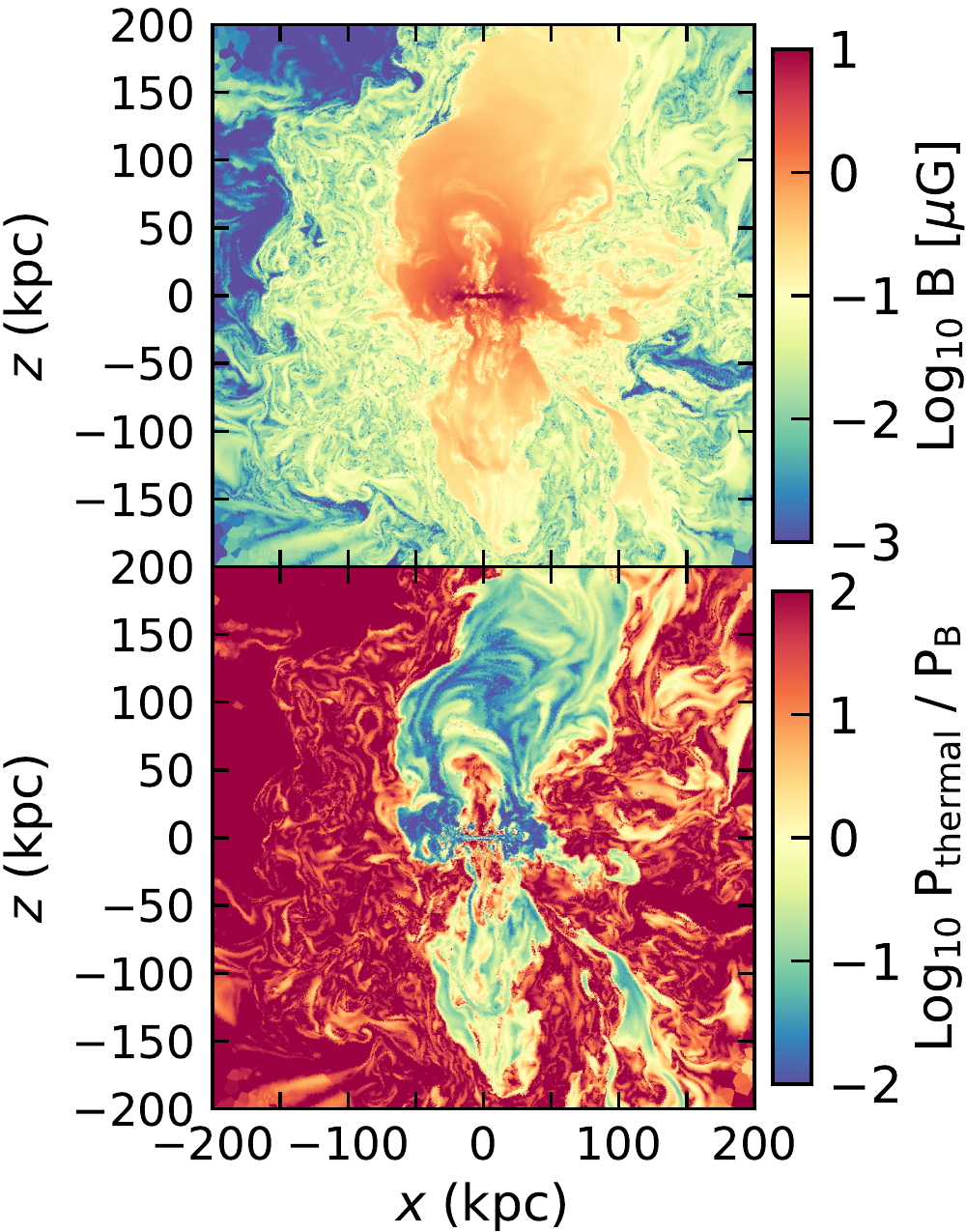}
\caption {\label{fig:imgBbeta} $400\times400$~kpc$^2$ images of the gas in and around an edge-on Milky Way-mass galaxy at $z=0$ in a simulation with magnetic fields. The top panel shows the magnetic field strength and the bottom panel shows the ratio $\beta$ between thermal pressure ($P_T=n k_\mathrm{B} T$) and magnetic pressure ($P_B=|\mathbf{B}|^2/8\pi$) in an infinitesimally thin slice through the centre of the edge-on galaxy. Even though thermal pressure dominates most of the volume, this is not the case everywhere. The magnetic field is much higher than average in the direction of the outflow (the polar direction). The combination of a higher field strength and a lower temperature means that the magnetic pressure can dominate over thermal pressure by 1 or 2 orders of magnitude.}
\end{figure}

The pressure shown in the middle panels is the sum of the thermal pressure and magnetic pressure. In the case of $B=0$ it is therefore equal to the thermal pressure. As seen before for the temperature, the pressure is higher when magnetic fields are included, the pressure is also smoother even in regions where the thermal pressure dominates (see Figure~\ref{fig:imgBbeta}). This pressure structure clearly connects to similar structure in the density and temperature images, which are also smoother when $B>0$. 

The metallicity distribution is strikingly different. Along the polar direction, the metallicity is high in both simulations. However, in the case where $B=0$, the metallicity is high in most of the halo volume. In the simulation with $B>0$, on the other hand, the metallicity in the regions away from the poles is much lower. This is likely due to reduced mixing of gas or due to outflows being more collimated along the polar axis when magnetic fields are included or a combination of both effects. 

The right-hand panels show the radial velocity, where inflow has negative and outflow has positive velocities. The radial velocity is much more time variable than the properties shown in the other panels, so the differences between top and bottom panels should not be overinterpreted. It is clear that the outflows are bipolar in both cases and correlate strongly with regions of high metallicity. The velocities are usually higher when $B=0$, although this is not the case for all simulation outputs between $z=0.3$ and $z=0$, and is quantified below in Figure~\ref{fig:prof}. The outflows are in general somewhat more confined to the polar axes when magnetic fields are included. This likely partially explains the difference in metal mixing.

The magnetic field (top panel) and $\beta = P_T / P_B$ (bottom panel) of the same simulation shown in the bottom panels of Figure~\ref{fig:imgBnoB} (`Auriga noAGN') are shown in Figure~\ref{fig:imgBbeta}. The magnetic field strength is clearly bipolar, with larger field strengths along the polar axes than in the plane of the disc, in qualitative agreement with \citet{Pakmor2020}. A turbulent dynamo increases the field strength in the galaxy after which galactic outflows magnetize the halo. This explains the strong correlation between areas of high magnetic field strength and high metallicity, since metals also have their origin inside the central galaxy and are expelled by outflows. Although the thermal pressure dominates in most of the halo volume, the magnetic pressure can dominate locally by up to two orders of magnitude, as can be seen in the bottom panel. As seen above in Figure~\ref{fig:imgBnoB}, the CGM properties are affected by the presence of $B$ fields in the entire halo, including in areas where the thermal pressure dominates.

\subsection{Properties of the central galaxy and its halo} \label{sec:prop}

\begin{figure}
\center
\includegraphics[scale=0.54]{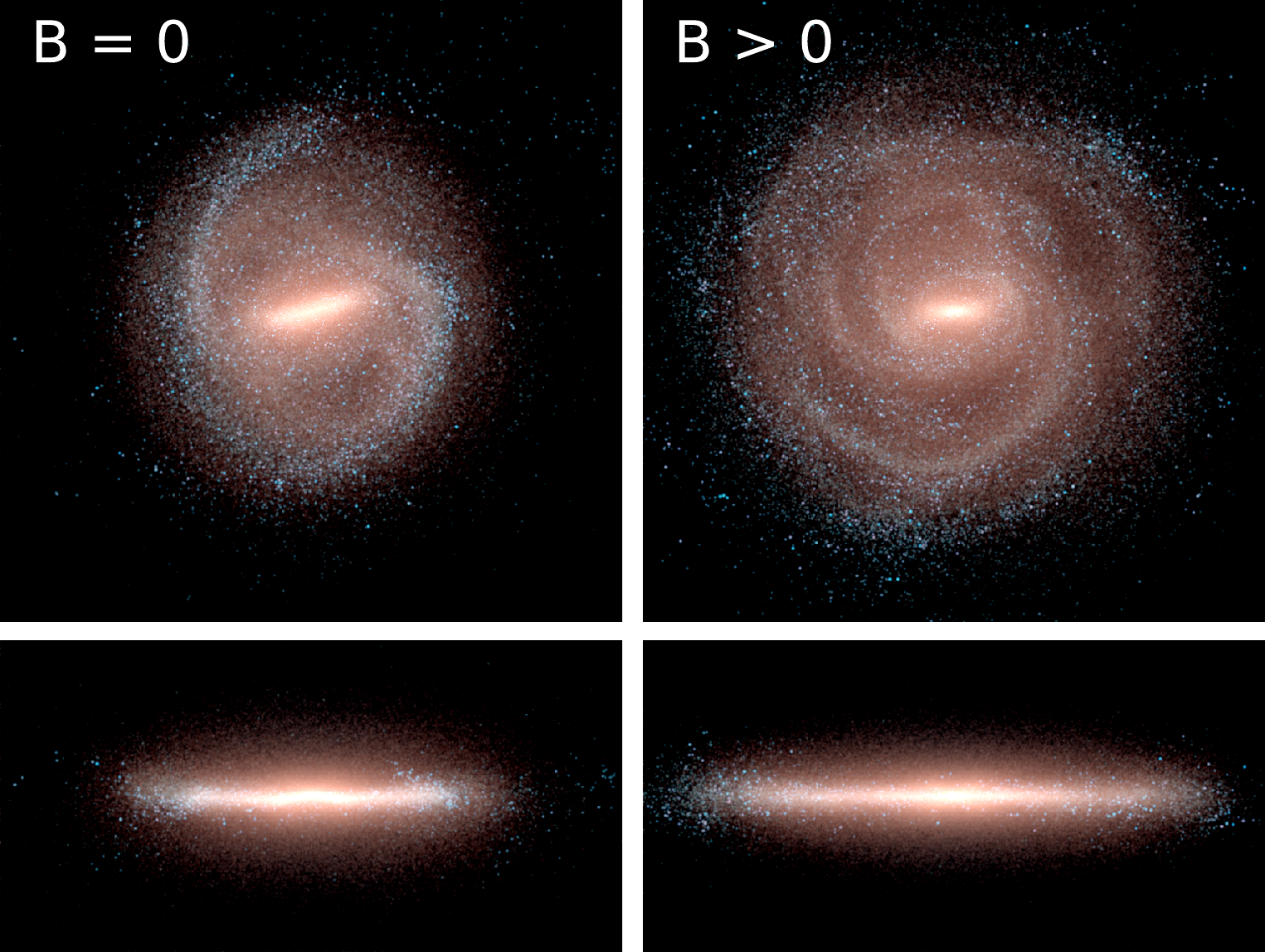}
\caption {\label{fig:stars} $50\times50$~kpc$^2$ face-on images and $50\times25$~kpc$^2$ edge-on images at $z=0$ in K-, B-, and U-band, which are shown by red, green and blue colour channels, respectively. Younger stars thus appear bluer. The left-hand (right-hand) panels show the `Auriga noAGN' simulation without (with) magnetic fields. The galaxy is dominated by a star-forming disc and also has a bulge comprised of older stars. The disc is more extended in the case of $B>0$ and the bar is more prominent when $B=0$.}
\end{figure} 

The central Milky Way-mass galaxy is shown in Figure~\ref{fig:stars} as a face-on and edge-on three-colour image (in K-, B-, and U-band) for model `Auriga noAGN' with $B=0$ (left-hand panels) and $B>0$ (right-hand panels). Both galaxies are disc dominated with a substantial bulge and have similar stellar mass and star formation rate. The galaxy in the simulation without magnetic fields has a stronger bar and a smaller disc than the galaxy with $B>0$. These differences are robust to changes in the galaxy formation model or CGM resolution. The two additional haloes with different initial conditions, presented in Appendix~\ref{sec:haloes}, also exhibit a smaller disc when $B=0$. However, the bar strength can both increase or decrease, so this seems to vary between different haloes. A larger sample of galaxies would be useful to verify whether magnetic fields always increase the extent of the stellar disc of Milky Way-mass galaxies. Our result is consistent with \citet{Pillepich2018}, who found that for $M_\mathrm{star}\gtrsim3\times10^{10}$~M$_{\astrosun}$ galaxy sizes increased when including magnetic fields, based on a larger sample of galaxies in a (25~Mpc)$^3$ simulation volume. It is also consistent with \citet{Whittingham2020} who found an even larger effect of magnetic fields on galaxy disc sizes after gas-rich major mergers. Overall, the changes in the stellar distribution are small enough that differences in the central galaxy are unlikely to have a significant impact on the properties of the CGM. Any variations in CGM properties due to magnetic fields are likely caused by the presence of $B$ fields in the CGM itself.

\begin{table*}
\begin{center}
\caption{\label{tab:prop} \small Properties of the galaxy and halo at $z=0$, unless otherwise stated, in our simulations with 1~kpc CGM refinement: galaxy formation model, inclusion of $B$ field, total stellar mass within 30~kpc from the centre ($M_\mathrm{star}$), mass of the central black hole ($M_\mathrm{BH}$), total ISM mass within 30~kpc from the centre ($M_\mathrm{ISM}$), total CGM mass (i.e.\ all non-star-forming gas within $R_\mathrm{vir}$; $M_\mathrm{CGM}$), the total amount of H~\textsc{i} in the CGM ($M_\mathrm{HI}$), star formation rate averaged over $z=0.3-0$ (SFR), and kinetic disc-to-total ratio based on $\kappa_\mathrm{rot}$ (D/T; see Equation~\ref{eqn:kappa}). Some of these properties are robust to the inclusion or exclusion of magnetic fields ($M_\mathrm{star}$, $M_\mathrm{ISM}$, SFR), whereas others are clearly affected ($M_\mathrm{BH}$, $M_\mathrm{CGM}$,  $M_\mathrm{HI}$, D/T).}
\vspace{-4mm}
\begin{tabular}[t]{lrrrrrrcr}
\hline \\[-3mm]
simulation & $B$ & log$_{10}$ $M_\mathrm{star}$ & log$_{10}$ $M_\mathrm{BH}$ &  log$_{10}$ $M_\mathrm{ISM}$ & log$_{10}$ $M_\mathrm{CGM}$ & log$_{10}$ $M_\mathrm{HI}$ & SFR                                    & D/T \\
name         & field         & [M$_{\astrosun}$] & [M$_{\astrosun}$]  &  [M$_{\astrosun}$]  & [M$_{\astrosun}$]   & [M$_{\astrosun}$]     & (M$_{\astrosun}$~yr$^{-1}$) & ($\kappa_\mathrm{rot}$)  \\
\hline \\[-4mm]                                                                                                                                       
\color{red} Auriga noAGN   & \color{red} no    & \color{red} ${10.85}$ & \color{red} --                   & \color{red} ${9.90}$  & \color{red} ${10.56}$ & \color{red} ${9.87}$ & \color{red} 6.6  & \color{red} 0.62   \\
\color{blue} Auriga noAGN & \color{blue} yes & \color{blue} ${10.90}$ & \color{blue} --                & \color{blue} ${9.94}$ & \color{blue} ${10.76}$ & \color{blue} ${10.38}$ & \color{blue} 6.1 & \color{blue} 0.71  \\[1mm]
\color{red} Auriga               & \color{red} no    & \color{red} ${10.69}$   & \color{red} ${7.33}$   & \color{red} ${9.76}$   & \color{red} ${10.71}$ & \color{red} ${9.43}$   & \color{red} 3.5  & \color{red} 0.58  \\
\color{blue} Auriga              & \color{blue} yes & \color{blue} ${10.68}$ & \color{blue} ${7.54}$ & \color{blue} ${9.55}$ & \color{blue} ${10.75}$ & \color{blue} ${10.09}$ & \color{blue} 2.9 & \color{blue} 0.74  \\[1mm]
\color{red} TNG                  & \color{red} no    & \color{red} ${10.58}$    & \color{red} ${7.53}$   & \color{red} ${9.68}$   & \color{red} ${10.61}$   & \color{red} ${9.89}$ & \color{red} 2.3  & \color{red} 0.66  \\
\color{blue} TNG                 & \color{blue} yes & \color{blue} ${10.63}$ & \color{blue} ${7.75}$  & \color{blue} ${9.82}$ & \color{blue} ${10.77}$ & \color{blue} ${10.28}$ & \color{blue} 3.6 & \color{blue} 0.78  \\[-1mm]
\hline
\end{tabular}
\end{center}
\end{table*}   

Several properties of the galaxy, its halo, and its central black hole are listed in Table~\ref{tab:prop} for all of our models, where magnetic fields are excluded (included) for the red (blue) entries. These simulations were all run with 1~kpc spatial refinement in the halo. Simulations with mass refinement only have very similar properties (see Table~\ref{tab:res} in Appendix~\ref{sec:test}). The first two columns indicate the galaxy formation model used and whether or not magnetic fields were included. The quantities listed are stellar mass (3rd column), black hole mass (4th column), ISM mass (5th column), CGM mass (6th column), H~\textsc{i} mass within the CGM (7th column), star formation rate (SFR; 8th column), and disc-to-total ratio (final column). The latter was determined based on kinematic properties by calculating the fraction of kinetic energy in ordered rotation \citep[e.g.][]{Sales2012}, calculated as follows: 
\begin{equation} \label{eqn:kappa}
  \kappa_\mathrm{rot}=K_\mathrm{rot} / K_\mathrm{tot}.
\end{equation}
$K_\mathrm{rot}$ is the kinetic energy in rotation around the angular momentum axis of the galaxy. This is calculated by summing over all star particles within 30~kpc from the centre of the galaxy using each particle's mass, $m$, its specific angular momentum in the direction of the angular momentum vector of the disc, $j_z$, and its distance to the angular momentum axis, $R_\mathrm{xy}$:
\begin{equation}
K_\mathrm{rot}=\sum \dfrac{1}{2} m \bigg( \dfrac{j_z}{R_\mathrm{xy}} \bigg)^2.
\end{equation}
$K_\mathrm{tot}$ is the total kinetic energy of the stars within 30~kpc from the centre, where $v$ is the total velocity of each particle:
\begin{equation}
K_\mathrm{tot}=\sum \dfrac{1}{2} m v^2.
\end{equation}

It can be seen from Tables~\ref{tab:prop} and~\ref{tab:res} that the stellar mass, ISM mass, and SFR do not change in a consistent way when adding $B$ fields, i.e.\ for some models they increase when $B>0$ and others they decrease. The same is true for the metallicity of the ISM. This is despite the fact that the star formation model \citep{Springel2003} was not recalibrated when including magnetic fields, which provide additional pressure in the ISM. This magnetic pressure is subdominant at high redshift, but becomes more important towards $z=0$ \citep{Pakmor2017}. We do not believe that not recalibrating the star formation model with regards to the additional pressure of the magnetic fields will strongly impact our results, because the differences in the stellar mass, ISM mass, and SFR due to the inclusion of magnetic fields are small and not systematic. 

Robust trends are present for black hole mass, CGM mass, H\,\textsc{i} mass, and disc-to-total ratio. Simulations with magnetic fields contain more massive central black holes, have more gas and more atomic hydrogen in their haloes, and their central galaxies are more disc-dominated. This is consistent with \citet{Pillepich2018} who simulated large volumes with the `TNG' model and compared simulations with and without magnetic fields. For Milky Way-mass galaxies they also find more massive black holes, higher gas fractions, and larger galaxy sizes. However, their stellar mass is more affected by the presence or absence of $B$ fields. This could be because they simulated a larger sample of galaxies and our single galaxy may be an outlier in terms of its stellar mass or it could be because the behaviour is different at lower mass resolution (by a factor 44). Note that most of the H~\textsc{i} is located in the central 30~kpc for all our simulations, which means that the H~\textsc{i} content of the CGM is dominated by a thick and extended non-star-forming disc surrounding the ISM of the galaxy.

\subsection{Galaxy formation model variations}

\begin{figure}
\center
\includegraphics[scale=0.6]{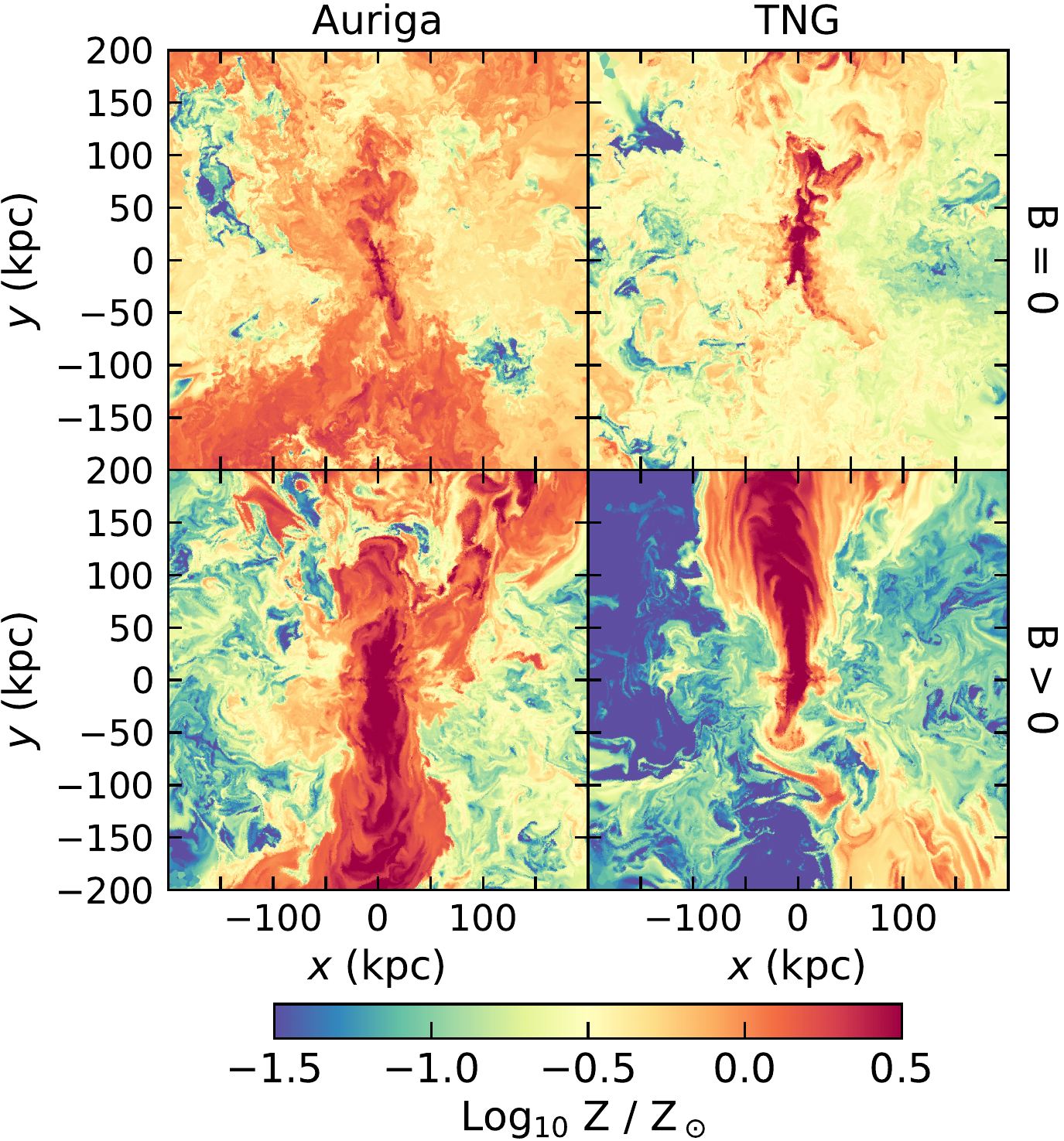}
\caption {\label{fig:imgZ} $400\times400$~kpc$^2$ images of the gas metallicity in and around a Milky Way-mass galaxy at $z=0$ in simulations without magnetic fields (top panels) and with magnetic fields (bottom panels), similar to the right-hand panels of Figure~\ref{fig:imgBnoB}. The left (right) panels show simulations run with the `Auriga' (`TNG') model. In both these models, which have very different AGN feedback implementations, the consequences of adding magnetic fields are similar to their effect in the `Auriga noAGN' simulation, shown in Figure~\ref{fig:imgBnoB}: the metals are more mixed throughout the CGM without magnetic fields and they remain more collimated in the direction of the outflows with magnetic fields.}
\end{figure} 

In order to check whether our results are sensitive to changes in the feedback model, we have repeated the experiment for the `Auriga' model and the `TNG' model, which differ by their implementation of AGN feedback, among other things (see Section~\ref{sec:sim}, \citealt{Grand2017}, \citealt{Weinberger2017}, and \citealt{Pillepich2018}). The `TNG' model also uses a different set of yields than the Auriga model and stronger stellar winds for low-mass galaxies (including for the progenitors of the Milky Way), which results in a lower metallicity overall. The resulting gas metallicity in a slice through the halo is shown in Figure~\ref{fig:imgZ}, which can be compared with model `Auriga noAGN' in the right-hand panels of Figure~\ref{fig:imgBnoB} (same colour scale). Although different in detail, the similarities of the impact of magnetic fields are striking. Comparing the simulations without $B$ fields (top panels) to those with (bottom panels), each simulation set clearly shows a more even spread of the metals throughout the halo when $B=0$ and a more confined high-metallicity region in the outflow direction when $B>0$, as seen before for the `Auriga noAGN' model. The metal-rich and metal-poor gas do not mix efficiently when magnetic fiels are included in the simulation, independent of feedback model. 

\begin{figure}
\center
\includegraphics[scale=0.6]{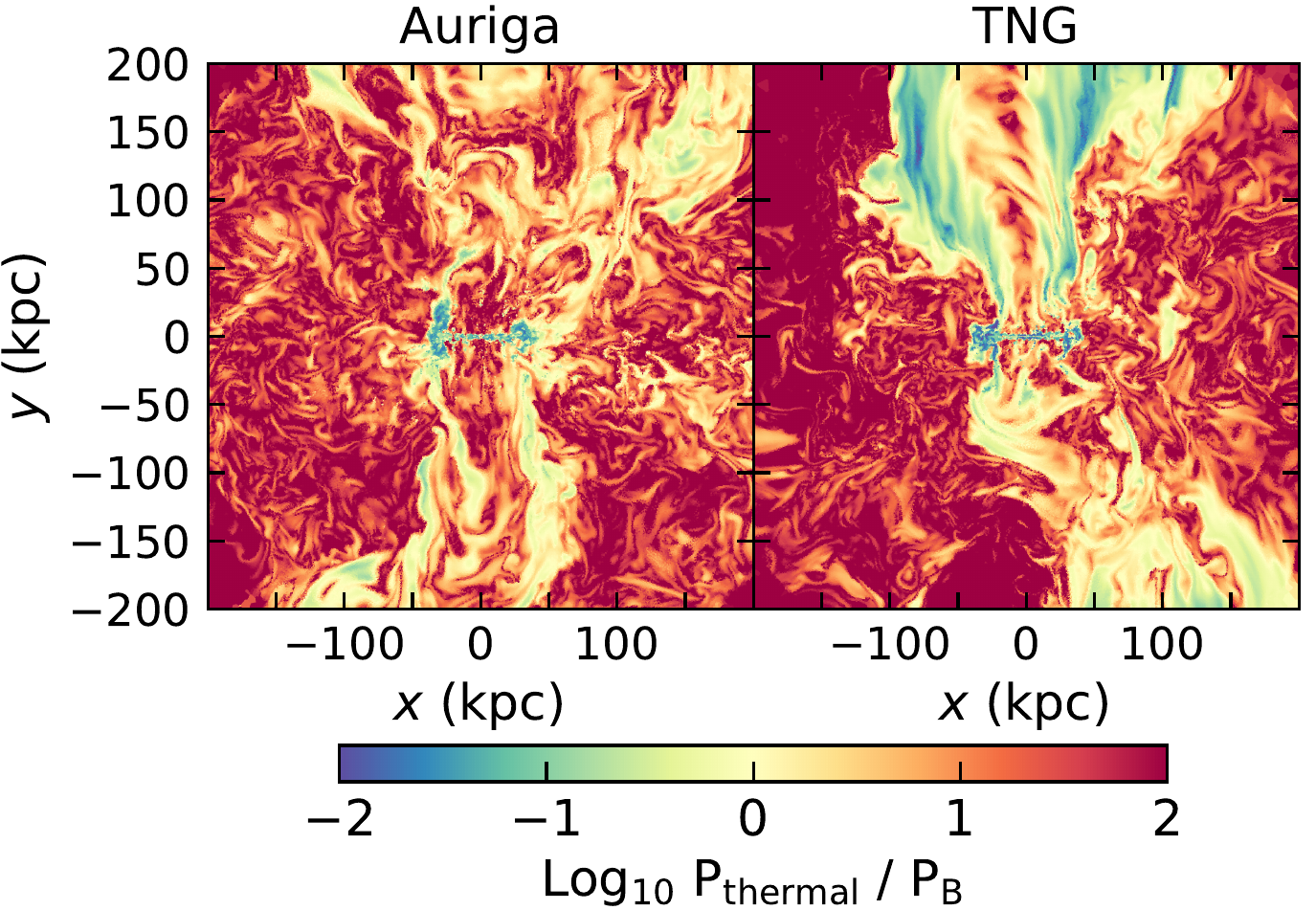}
\caption {\label{fig:imgbeta} $400\times400$~kpc$^2$ images of $\beta$, the ratio between thermal pressure and magnetic pressure, in an infinitesimally thin slice through the centre of the edge-on galaxy for model `Auriga (left-hand panel) and model `TNG' (right-hand panel), similar to the bottom panel in Figure~\ref{fig:imgBbeta}. For all models, most of the volume is dominated by thermal pressure. However, some of the gas in the centre (the dense, extended gas disc around the galaxy) and in the outflow direction is dominated by magnetic pressure. The area in which $\beta\ll1$ is somewhat reduced in these models with AGN feedback as compared to the simulation without AGN feedback (see Figure~\ref{fig:imgBbeta}).}
\end{figure} 

The ratio between thermal and magnetic pressure, or $\beta$, in the CGM is shown in Figure~\ref{fig:imgbeta} for models `Auriga' and `TNG'. The total area in the CGM where magnetic fields strongly dominate the pressure is reduced when compared to model `Auriga noAGN' in Figure~\ref{fig:imgBbeta}. The `Auriga' model looks the most discrepant with somewhat less ordered magnetic field structure, likely because its AGN radio-mode feedback is initialized in random locations in the halo rather than in the centre of the galaxy. However, there are many similarities between models as well. The extended, thick disc around the galaxy is cool and dense and has a large magnetic field strength, making it strongly magnetic pressure dominated ($\beta\ll1$). Additionally, a biconical structure is visible, similar to that seen for model `Auriga noAGN'. Both the magnetic field strength and the metallicity are largest in the outflow direction. Some of this high-metallicity gas cools to lower temperatures and the low thermal pressure in combination with the high magnetic pressure result in $\beta\ll1$ in these regions.

\subsection{Quantifying the effect of magnetic fields} \label{sec:prof}

\begin{figure}
\center
\includegraphics[scale=0.6]{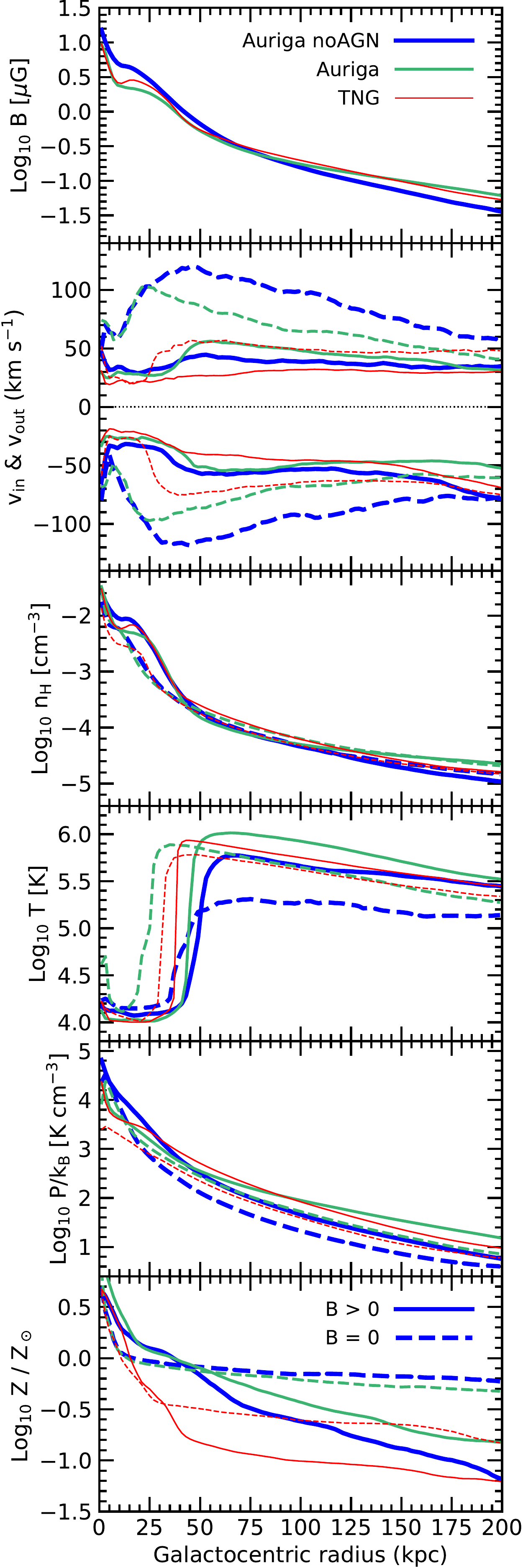}
\caption {\label{fig:prof} CGM properties as a function of galactocentric radius. Gas in the ISM is excluded. (See separate caption for more information.)}
\end{figure} 

\begin{figure}
\contcaption{The thick, blue curves show results from the `Auriga' model without AGN feedback, the green curves show those from the `Auriga' model, and the thin, red curves show those based on the `TNG' model. Solid (dashed) curves represent simulations with (without) magnetic fields. We average over all snapshots between $z=0.3$ and $z=0$ and exclude gas in the ISM and gas associated with satellites. The first panel shows the volume-weighted root-mean-square magnetic field strength. The second panel shows the mass-weighted median radial velocity separately for inflowing ($v_\mathrm{rad}<0$) and outflowing ($v_\mathrm{rad}>0$) gas. The black, dotted curve shows the dividing line for inflow and outflow at $v_\mathrm{rad}=0$. The average hydrogen number density is shown in the third panel. The fourth panel shows the mass-weighted median temperature and the fifth panel the volume-weighted median pressure (thermal + magnetic), divided by the Boltzmann constant. The mass-weighted median metallicity is shown in the final panel. The presence of magnetic fields reduces the inflow velocity (making it less negative) and outflow velocity, enhances the density in the centre, decreases the temperature outside $\approx50$~kpc, increases the pressure, and decreases the amount of mixing in the CGM resulting in steeper metallicity profiles.}
\end{figure}

The difference between simulations with and without magnetic fields is quantified in Figure~\ref{fig:prof}, which shows several CGM properties as a function of galactocentric radius, $R_\mathrm{GC}$. Gas in the ISM and gas associated with satellites has been excluded from these profiles. From top to bottom, the panels show magnetic field strength, radial velocity for inflow (negative velocities) and outflow (positive velocities) separately, hydrogen number density, temperature, total pressure (thermal plus magnetic), and metallicity. Simulations with $B=0$ are shown by the dashed curves and for $B>0$ as solid curves for models `Auriga noAGN' (thick, blue curves), `Auriga' (green curves), and `TNG' (thin, red curves). To limit the impact of stochastic processes, we show properties averaged over all 22 simulation outputs between $z=0.3$ and $z=0$, but this choice does not affect our conclusions.

Differences caused by the presense or absense of $B$ fields are generally largest in the `Auriga noAGN' model and smaller in the two models with AGN feedback, which could indicate that additional feedback reduces some of the effects of magnetic fields on the CGM. Even though the magnitude of the effect of magnetic fields varies between galaxy formation models, the simulations show the same qualitative behaviour. Magnetic fields are clearly important for a wide range of CGM properties in all of our simulations. 

The strength of the magnetic field is reasonably similar in our three models. The difference is largest, up to 0.3~dex, in the central 30~kpc and in the halo outskirts. This means that the magnetic pressure ($P_B=|\mathbf{B}|^2/8\pi$) is up to 0.6~dex different in our models. However, the quantitative change in properties with or without $B$ does depend on the feedback model. The simulation with the lowest magnetic fields at large $R_\mathrm{GC}$ (`Auriga noAGN') is affected more strongly in some properties, e.g.\ outflow velocity and temperature, than the other models. Although the strength of the field must be important for how much the gas flows in the CGM are affected, strong feedback can limit their influence somewhat. It is also important to note that the strength of the magnetic field is much larger along the polar axes than in the disc plane (see Figure~\ref{fig:imgBbeta}). The regions where the field is relatively weak dominate the halo volume and therefore dominate our spherical averages in Figure~\ref{fig:prof}. 

The median radial velocity is negative at all radii: all of our haloes are dominated by inflowing gas. Instead of showing this, we split the gas into inflowing and outflowing gas to show both separately in the second panel of Figure~\ref{fig:prof}. The lower (higher) curves show the median for the gas with $v_\mathrm{rad}<0$ ($v_\mathrm{rad}>0$). For all our galaxy formation models and at all radii, the gas is inflowing and outflowing faster without magnetic fields. $B$ fields thus slow down the flow of gas in both directions. 

\begin{figure}
\center
\includegraphics[scale=0.66]{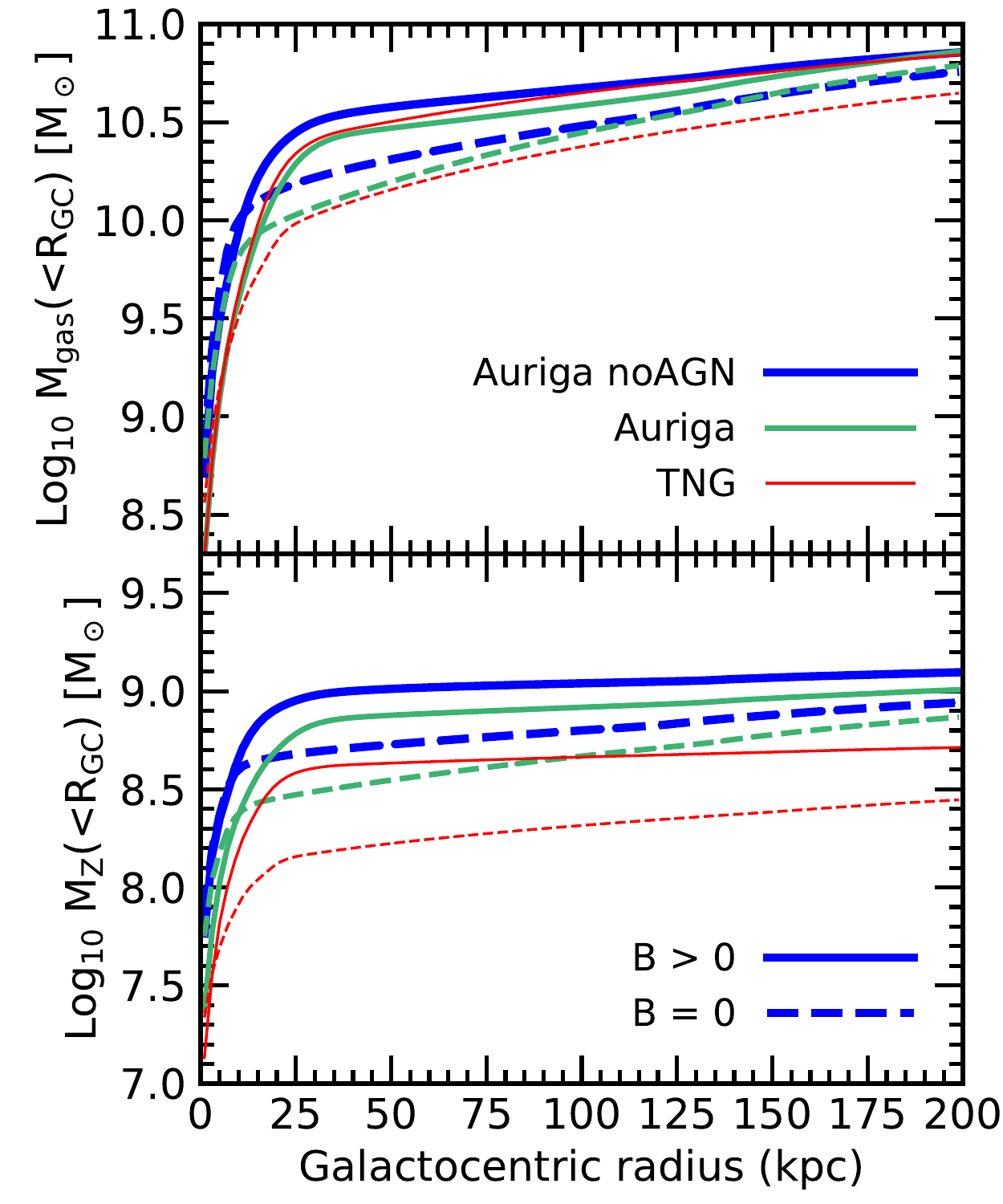}
\caption {\label{fig:cum} Cumulative mass and metal mass profiles of all the gas (ISM and CGM) as a function of galactocentric radius averaged over $z=0.3-0$. The baryon fraction and metal fraction of the halo are reduced in the presence of magnetic fields. The largest difference in mass within $R_\mathrm{GC}$ is visible for the innermost part of the CGM, for $R_\mathrm{GC}=25-50$~kpc. The difference decreases towards larger radii. Gas and metals are transported to large distances more efficiently without magnetic fields.}
\end{figure}

The density profiles are fairly similar and at $R_\mathrm{GC}>50$~kpc there is no consistent trend in the different models when including magnetic fields. Closer to the galaxy, all models show higher densties when $B>0$. Note that our radial profiles are averaged over spherical shells, whereas the gas at these lower radii is distributed in a thick disc (see Figure~\ref{fig:imgBnoB}). This dense, cool, non-star-forming gas disc is more massive when $B$ fields are present. Cumulative mass profiles are shown and discussed below (see Figure~\ref{fig:cum}). 

Because the cooling curve of metal-rich gas peaks at intermediate temperatures $T=10^{5-5.5}$~K, the CGM is multiphase with a large fraction of the gas at $T\approx10^4$~K (below which cooling is inefficient) and at $T\approx10^6$~K (the virial temperature of a Milky Way-mass halo). In the inner CGM, most of the gas is cool, whereas at larger radii, most of the gas is hot in all our simulations. There is relatively little gas at intermediate temperatures, which leads to a sharp increase at $R_\mathrm{GC}=20-50$~kpc in the mass-weighted median temperature profiles. In all our models, the temperature of the halo gas is higher outside $R\approx50$~kpc when including $B$ fields. How much the temperature is enhanced depends on the galaxy formation model. 

As required by hydrostatic equilibrium, the gas pressure decreases with galactocentric radius. However, in some of the simulations, the pressure briefly increases at small radius, likely caused by a cooling flow and thus a departure from hydrostatic equilibrium. Overall, the thermal pressure dominates, but at small radii ($R_\mathrm{GC}\approx25$~kpc) the magnetic pressure can be dominant over the thermal pressure, whilst the dynamics are dominated by rotation and kinetic energy. In these inner regions, the total pressure is higher with magnetic fields, even though the thermal pressure is actually lower (not shown). At $R_\mathrm{GC}\gtrsim50$~kpc, $P_\mathrm{thermal}\gg P_\mathrm{B}$, but the total pressure is nonetheless higher when $B>0$. Because the temperature in the CGM is generally higher in the presence of magnetic fields, its pressure is also somewhat higher. In the inner 40~kpc, the average density is higher with $B$, again leading to a higher pressure. Note that the temperature shown is mass-weighted and its value therefore does not directly translate directly to the volume-weighted pressure.  

The differences in CGM metallicity for simulations with or without magnetic fields are substantial. In the inner regions, the median metallicity is higher when $B>0$, whereas at $R_\mathrm{GC}\gtrsim50$~kpc, it is significantly lower. This is likely caused by a combination of less efficient mixing and more collimated metal-rich outflows, which means that a larger fraction of the CGM remains metal poor. We discuss this further below.

Cumulative profiles of the gas mass and metal mass in the gas-phase are shown in Figure~\ref{fig:cum}. This includes both the ISM and the CGM. At radii where the CGM dominates ($R_\mathrm{GC}\gtrsim25$~kpc), the baryon fraction and the metal fraction are reduced substantially when $B>0$, by up to 0.5~dex.  The difference is largest in the inner CGM, at $R_\mathrm{GC}=25-50$~kpc, and decreases towards larger radii. Outflows move gas and metals to larger distances without magnetic fields. Given that the stellar mass is similar with and without $B$ fields, this is unlikely caused by a change in feedback efficiency. Moreover, in the simulations with AGN feedback, the black hole mass and therefore the amount of energy put into winds driven by AGN is actually higher with $B$ fields. If this were important, we would expect that simulations with $B>0$ would have lower gas and metal fractions as more material is pushed out, but we find the opposite result. From this we can therefore confidently conclude that magnetic fields in the CGM change the circumgalactic gas flows, work to reduce the mass and metal outflow rate, and enhance the baryon fraction within the virial radius.

The metal mass follows the same trend as the gas mass, although the difference between simulations with or without $B$ fields is somewhat larger in the outer halo for the metal mass than for the total gas mass. This is because metals in the CGM were produced in the ISM and subsequently stripped from a galaxy or ejected in an outflow (with large $B$). However, gas can be stripped or ejected from galaxies (with large $B$) as well as accreted from the IGM (with small $B$). Accretion from the IGM is especially important at large radii. The metals are thus more closely associated with highly magnetized regions than the gas in general. In the halo outskirts, the total metal mass is therefore affected somewhat more strongly by magnetic fields than the total gas mass.

\subsection{Angular dependence, scatter, and mixing}

\begin{figure}
\center
\includegraphics[scale=0.66]{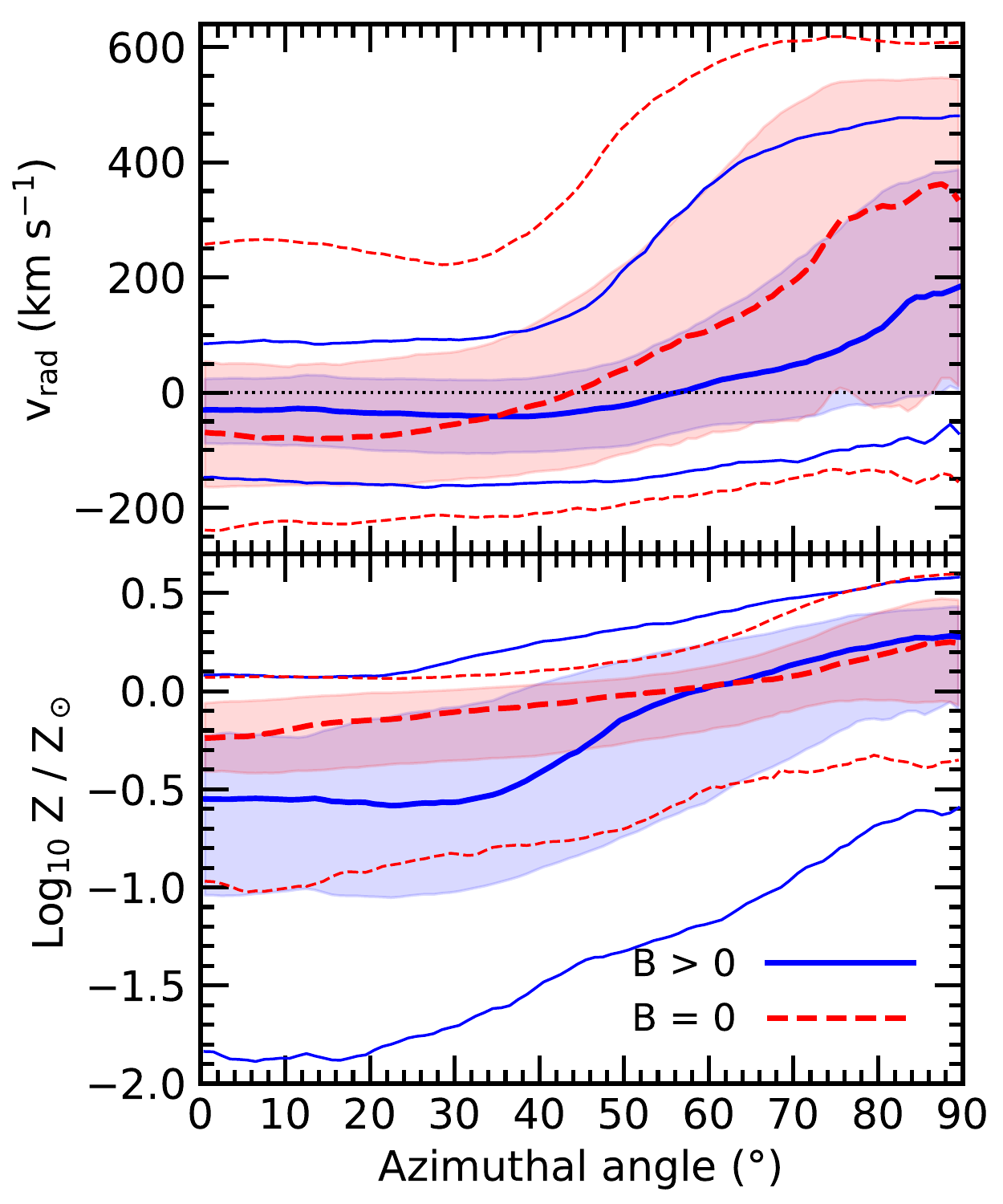}
\caption {\label{fig:angle} Angular dependence of the radial velocity and metallicity for the `Auriga noAGN' simulations with magnetic fields (solid, blue curves) and without magnetic fields (dashed, red curves) for the CGM gas with $R_\mathrm{GC}=50-100$~kpc averaged over $z=0.3-0$. Gas associated with satellites has been excluded. The plane of the disc, or major axis, lies at an angle of 0\textdegree~and the polar direction, or minor axis lies at 90\textdegree. Thick curves show the median and thin curves the $2\sigma$ scatter, while the shaded regions indicate the $1\sigma$ scatter. Both simulations show a clear angular dependence: the majority of the gas is inflowing in the disc plane and outflowing in the polar direction and the metallicity increases towards the polar axis. The scatter in radial velocity is \emph{smaller}, whereas the scatter in metallicity is \emph{larger} with magnetic fields. Less variation in velocity means that the gas flows are more collimated and more variation in metallicity means that the gas mixes less efficiently.}
\end{figure} 

The properties of the CGM in the plane of the disc are quite different from those along the polar direction, which is clearly visible in Figure~\ref{fig:imgBnoB} and which has been quantified for a statistical sample of simulated galaxies by \citet{Peroux2020}. We therefore study the angular dependence of the radial velocity and the metallicity and their scatter for gas between $R_\mathrm{GC}=50$ and 100~kpc in Figure~\ref{fig:angle}, where we define the azimuthal angle to be 0\textdegree~in the disc plane (major axis) and 90\textdegree~in the direction of the angular momentum vector (minor axis). We again average over all 22 snapshots between $z=0.3$ and $z=0$ to reduce the effect of variability in the outflows. We exclude gas associated with satellites. The thick curves show the median for a simulation without (dashed, red curves) and with (solid, blue curves) magnetic fields. The 16th and 84th percentile ($1\sigma$ scatter) ranges are shown by the shaded areas and the thin curves show the 98th and 2nd percentiles ($2\sigma$ scatter). Note that we calculated this in 3D (using segments of cones) and the angular dependence is likely somewhat weaker in 2D due to projection effects. Because the binsize is fixed at 2\textdegree, the number of resolution elements included in the bin decreases towards larger azimuthal angles (because the cone subtended gets narrower). Bins close to the polar axis are therefore noisier than those near the disc plane. 

The radial velocity increases from the disc plane, where the gas is predominantly inflowing, to the angular momentum axis, where it is mostly outflowing. This trend exists for all our simulations, independent of feedback model or magnetic fields, in agreement with \citet{Peroux2020}. The scatter, however, does depend on the inclusion of magnetic fields and is smaller when $B>0$. Simulations without magnetic fields have both stronger inflows and stronger outflows, as also shown in Figure~\ref{fig:prof}. Magnetic fields slow down the circumgalactic gas flows and also keep inflowing and outflowing regions more collimated, as evidenced by the reduced scatter.

The metallicity also increases from the disc plane to the polar axes, as was the case for the radial velocity, again in agreement with \citet{Peroux2020}. The scatter, however, is substantially larger when including magnetic fields. The distribution is not Gaussian, since the scatter to low metallicities is much larger than to high $Z$. Patches of gas with low metallicity are much more common when magnetic fields are present. This is likely due to inflowing gas and outflowing gas mixing less efficiently in the presence of $B$ fields. The reason for reduced mixing may be that the presence of magnetic fields suppresses hydrodynamic instabilities in the CGM and thus reduces mixing, as also found in small-scale simulations \citep{Berlok2019, Sparre2020}. However, this may be surprising, because the field strength is relatively low in the plane of the disc and the thermal pressure dominates strongly over the magnetic pressure (see Figure~\ref{fig:imgBbeta}). It is also possible that the metallicity difference is seeded at smaller scales near the galactic disc. In this region, the magnetic field is strong and able to confine outflows into cones along the polar axes and to inhibit outflows in other directions. Without magnetic fields, outflows could have a larger probability of escaping at an angle away from the poles and reach further into the halo. Follow-up work is needed to determine whether changes in direction of the outflows or suppression of hydrodynamic instabilities in the CGM is the dominant process responsible for reducing the amount of metal mixing. 

Observations of CGM absorption line systems have found a wide distribution of metallicities, both at low and high redshift \citep[e.g.][]{Fumagalli2016, Prochaska2017, Lehner2019, Zahedy2019}. This appears to be in better agreement with our simulation that includes magnetic fields, although a more detailed comparison -- where the observational sample is matched -- is necessary to draw any strong conclusions. An azimuthal dependence of various gas properties and their scatter could potentially be explored with observations of absorption lines in quasar spectra when the orientation of the galaxy nearest to the absorbing cloud is known \citep[e.g.][]{Kacprzak2012, Kacprzak2015, Martin2019, Schroetter2019}. Such a comparison could help determine how collimated the gas flows are and how well the CGM is mixed. However, even though all our simulations show the same trend with azimuthal angle, there are some quantitative differences depending on galaxy formation model, which will need to be taken into account when using simulations to help interpret observational data. Furthermore, a larger sample of simulated galaxies will be necessary to study the level of halo-to-halo variation. Finally, the results presented in this paper include all the gas, whereas absoption-line observations are sensitive to gas in a particular ionization state. A more detailed study will thus be needed for a fair comparison to observations, which we leave for future work.

\section{Discussion and Conclusions} \label{sec:concl}

We studied the effect of magnetic fields on the circumgalactic medium in cosmological, magnetohydrodynamical simulations of a Milky Way-mass galaxy using the moving mesh code \textsc{arepo}. This work is part of the SURGE project. We used three variations of the galaxy formation model, i.e.\ the Auriga model without AGN feedback, the Auriga model, and the TNG model and added extra spatial refinement to the CGM, so that its resolution was 1~kpc or better. Each simulation was run with the standard seed magnetic field (i.e.\ $\approx10^{-10}$~G initialized at $z=127$) and repeated with the seed field set to zero (i.e.\ without magnetic fields) in order to understand the importance of magnetic fields for setting the properties of the CGM. The main conclusions from this work are as follows.

\begin{enumerate}

\item Certain bulk properties of the central galaxies are unaffected by the inclusion of magnetic fields: the stellar masses, ISM masses, and SFRs are similar when $B=0$ and $B>0$ and there is no general trend. However, the galaxy is more disc dominated and the central black hole is more massive when $B$ fields are included. This is the case for all of our models and for both standard simulations and simulations with CGM refinement. The extended H\,\textsc{i} disc surrounding the galaxy is much more massive in the simulations with $B>0$. 
  
\item The properties and structure of the CGM are affected by the presence or absence of magnetic fields, even in regions where the thermal pressure dominates over the magnetic pressure. Although there are quantitative differences for different galaxy formation models, all our simulations show the same qualitative effect. Magnetic fields increase the density in the inner CGM and the temperature in the outer CGM. The structure of the CGM also changes, with more elongated filaments when $B>0$. Magnetic fields increase the total pressure and reduce pressure variations, resulting in a smoother CGM structure. $B$ fields furthermore decrease the amount of mixing between metal-poor and metal-rich gas. This is likely because the bipolar outflows are more collimated when magnetic fields are included and because the radial velocity of both inflowing and outflowing gas is substantially reduced. 

\item The fact that the gas flows are slower when $B>0$ also has important implications for the amount of mass and metals present within the CGM. The presense of magnetic fields increases both the gas fraction and the metal mass fraction inside the halo. When $B$ fields are excluded, the outflows more easily escape the halo and push more metal-rich material out beyond the virial radius. 

\item The radial velocity and metallicity have a clear angular dependence. Most of the gas is metal-rich and outflowing in the angular momentum direction (along the minor axis) and metal-poor and inflowing in the plane of the disc (along the major axis). This is the case for both with and without $B$ fields, although the angular dependence of the metallicity is somewhat stronger when $B>0$. The scatter in CGM properties is also affected by magnetic fields. Their inclusion decreases the scatter in radial velocity and increases the scatter in metallicity, which indicates that the gas flows are more collimated and that the CGM less well mixed. 
  
\item We repeated our simulations without additional CGM refinement and found very similar results (see Appendix~\ref{sec:test}). We therefore conclude that our conclusions are robust to changes in resolution. The most noticeable change due to resolution is seen for the metallicity of the CGM, which is more affected by the presence of magnetic fields at higher resolution. This is likely caused by reduced numerical mixing at higher resolution. It is therefore possible that even our high-resolution simulations are underestimating the impact of magnetic fields on metal mixing in the CGM. 

\end{enumerate}

Because we see the same effect in all of our simulations, i.e.\ in 8 pairs of simulations in total, independent of resolution (see Appendix~\ref{sec:test}) or exact initial conditions (see Appendix~\ref{sec:haloes}) and because the results shown are averaged over all 22 simulation outputs between $z=0.3$ and $z=0$, we are confident that none of our conclusions are caused by random stochastic processes, such as the specific timings of outflows or galaxy mergers. We wish to emphasize, however, that we have only studied haloes with $M_\mathrm{halo}=10^{12}$~M$_{\astrosun}$ at low redshift and more work is needed to determine whether or not magnetic fields affect the CGM for different halo masses and at higher redshifts.

Although our simulations are well-converged within the kpc-scale resolution regime of cosmological simulations, any structure (e.g.\ shocks, clumps, turbulence, and vorticity) on scales below our resolution cannot be resolved. It is therefore possible that our simulations underestimate the amplification of the magnetic field due to small-scale processes \citep[e.g.][]{Inoue2013, Ji2016}. Additionally, because we did not include cosmic rays, certain instabilities that could lead to further magnetic field amplification are also not included \citep[e.g.][]{Bell2004, Beresnyak2009}. Larger magnetic field strengths could potentially increase their effect on the CGM beyond what we see in our cosmological simulations.

Idealized studies can reach orders of magnitude higher resolution and resolve much smaller clouds or filaments and thinner interfaces between the hot and cool gas than the cosmological simulations presented here. Such simulations find that magnetic fields can enhance the survival of cool, relatively dense gas, substantially change its morphology, and reduce its mixing with the surrounding hot, diffuse gas \citep[e.g.][]{Dursi2008, McCourt2015, Ji2018, Berlok2019, Gronke2020, Liang2020}. Increased magnetic tension as magnetic field lines are swept up has been shown to result in a decrease of the relative velocities of clouds and their background medium \citep[e.g.][]{Dursi2008, McCourt2015, Gronke2020}. These results are at least in qualitative agreement with our findings of larger cool gas masses, slower velocities, and less mixing in our Milky Way-mass halo when magnetic fields are included. 

Using non-ideal MHD instead of ideal MHD would likely not have any effect at the current resolution of our cosmological simulations, where numerical dissipation of the magnetic fields dominates. Regardless, the difference between including or excluding magnetic fields are likely substantially larger than the changes expected from including non-ideal MHD terms. Given the sizable variation seen between our three different galaxy formation models, we believe that uncertainties in CGM properties due to uncertainties in feedback modelling exceed those from not including non-ideal MHD. We reiterate that, although the level of change due to magnetic fields varies between galaxy formation models, each of our models exhibits qualitatively similar and non-negligible effects of including $B$ fields.

We have shown that magnetic fields change the flow of circumgalactic gas and its physical properties, such as its density, temperature, and pressure. They are also important for other physical processes not yet included in our simulation suite. Examples of these processes are thermal conduction \citep[e.g.][]{Sharma2010, Armillotta2016, Bruggen2016, Li2020} and cosmic ray feedback \citep[e.g.][]{Simpson2016, Butsky2018, Chan2019, Wiener2019}, which primarily act along magnetic field lines. These additional processes may also affect the CGM as well as its embedded galaxies and will be explored in future work. Here, we have shown that magnetic fields have a significant impact on the CGM and therefore should be included in simulations of galaxy formation.

\section*{Acknowledgements}

We would like to thank the IllustrisTNG team for the use of their galaxy formation model and the referee and editor for their helpful comments that improved our manuscript. 
FvdV is supported by a Royal Society University Research Fellowship and was also supported by the Deutsche Forschungsgemeinschaft through project SP 709/5-1.
FAG acknowledges financial support from CONICYT through the project FONDECYT Regular Nr. 1181264, and funding from the Max Planck Society through a Partner Group grant.
FM is supported by the program ``Rita Levi Montalcini'' of the Italian MIUR.
Our simulations of halo~6 were performed on computing resources provided by the Max Planck Computing and Data Facility in Garching. 
For the simulations of halo~12 and~L8, the authors gratefully acknowledge the Gauss Centre for Supercomputing e.V.\ (\url{https://www.gauss-centre.eu}) for funding this project by providing computing time on the GCS Supercomputer SuperMUC-NG at Leibniz Supercomputing Centre (\url{https://www.lrz.de}).

\section*{Data availability}

Data available on request.

\bibliographystyle{mnras}
\bibliography{Bfields}

\begin{thebibliography}{}
\makeatletter
\relax
\def\mn@urlcharsother{\let\do\@makeother \do\$\do\&\do\#\do\^\do\_\do\%\do\~}
\def\mn@doi{\begingroup\mn@urlcharsother \@ifnextchar [ {\mn@doi@}
  {\mn@doi@[]}}
\def\mn@doi@[#1]#2{\def\@tempa{#1}\ifx\@tempa\@empty \href
  {http://dx.doi.org/#2} {doi:#2}\else \href {http://dx.doi.org/#2} {#1}\fi
  \endgroup}
\def\mn@eprint#1#2{\mn@eprint@#1:#2::\@nil}
\def\mn@eprint@arXiv#1{\href {http://arxiv.org/abs/#1} {{\tt arXiv:#1}}}
\def\mn@eprint@dblp#1{\href {http://dblp.uni-trier.de/rec/bibtex/#1.xml}
  {dblp:#1}}
\def\mn@eprint@#1:#2:#3:#4\@nil{\def\@tempa {#1}\def\@tempb {#2}\def\@tempc
  {#3}\ifx \@tempc \@empty \let \@tempc \@tempb \let \@tempb \@tempa \fi \ifx
  \@tempb \@empty \def\@tempb {arXiv}\fi \@ifundefined
  {mn@eprint@\@tempb}{\@tempb:\@tempc}{\expandafter \expandafter \csname
  mn@eprint@\@tempb\endcsname \expandafter{\@tempc}}}

\bibitem[\protect\citeauthoryear{{Armillotta}, {Fraternali}  \&
  {Marinacci}}{{Armillotta} et~al.}{2016}]{Armillotta2016}
{Armillotta} L.,  {Fraternali} F.,   {Marinacci} F.,  2016, \mnras, \href
  {https://ui.adsabs.harvard.edu/abs/2016MNRAS.462.4157A} {462, 4157}

\bibitem[\protect\citeauthoryear{{Barnes} et~al.,}{{Barnes}
  et~al.}{2019}]{Barnes2019}
{Barnes} D.~J.,  et~al., 2019, \mnras, \href
  {https://ui.adsabs.harvard.edu/abs/2019MNRAS.488.3003B} {488, 3003}

\bibitem[\protect\citeauthoryear{{Beck} \& {Wielebinski}}{{Beck} \&
  {Wielebinski}}{2013}]{Beck2013}
{Beck} R.,  {Wielebinski} R.,  2013, in {Oswalt} T.~D.,  {Gilmore} G.,  eds,
  Planets, Stars and Stellar Systems Vol. 5, Galactic Structure and Stellar
  Populations. Springer Science+Business Media Dordrecht, p.~641

\bibitem[\protect\citeauthoryear{{Bell}}{{Bell}}{2004}]{Bell2004}
{Bell} A.~R.,  2004, \mnras, \href
  {https://ui.adsabs.harvard.edu/abs/2004MNRAS.353..550B} {353, 550}

\bibitem[\protect\citeauthoryear{{Beresnyak}, {Jones}  \&
  {Lazarian}}{{Beresnyak} et~al.}{2009}]{Beresnyak2009}
{Beresnyak} A.,  {Jones} T.~W.,   {Lazarian} A.,  2009, \apj, \href
  {https://ui.adsabs.harvard.edu/abs/2009ApJ...707.1541B} {707, 1541}

\bibitem[\protect\citeauthoryear{{Berlok} \& {Pfrommer}}{{Berlok} \&
  {Pfrommer}}{2019}]{Berlok2019}
{Berlok} T.,  {Pfrommer} C.,  2019, \mnras, \href
  {https://ui.adsabs.harvard.edu/abs/2019MNRAS.489.3368B} {489, 3368}

\bibitem[\protect\citeauthoryear{{Br{\"u}ggen} \& {Scannapieco}}{{Br{\"u}ggen}
  \& {Scannapieco}}{2016}]{Bruggen2016}
{Br{\"u}ggen} M.,  {Scannapieco} E.,  2016, \apj, \href
  {https://ui.adsabs.harvard.edu/abs/2016ApJ...822...31B} {822, 31}

\bibitem[\protect\citeauthoryear{{Buck}, {Pfrommer}, {Pakmor}, {Grand}  \&
  {Springel}}{{Buck} et~al.}{2020}]{Buck2020}
{Buck} T.,  {Pfrommer} C.,  {Pakmor} R.,  {Grand} R. J.~J.,   {Springel} V.,
  2020, \mnras, \href {https://ui.adsabs.harvard.edu/abs/2020MNRAS.497.1712B}
  {497, 1712}

\bibitem[\protect\citeauthoryear{{Butsky} \& {Quinn}}{{Butsky} \&
  {Quinn}}{2018}]{Butsky2018}
{Butsky} I.~S.,  {Quinn} T.~R.,  2018, \apj, \href
  {https://ui.adsabs.harvard.edu/abs/2018ApJ...868..108B} {868, 108}

\bibitem[\protect\citeauthoryear{{Chan}, {Kere{\v{s}}}, {Hopkins}, {Quataert},
  {Su}, {Hayward}  \& {Faucher-Gigu{\`e}re}}{{Chan} et~al.}{2019}]{Chan2019}
{Chan} T.~K.,  {Kere{\v{s}}} D.,  {Hopkins} P.~F.,  {Quataert} E.,  {Su} K.~Y.,
   {Hayward} C.~C.,   {Faucher-Gigu{\`e}re} C.~A.,  2019, \mnras, \href
  {https://ui.adsabs.harvard.edu/abs/2019MNRAS.488.3716C} {488, 3716}

\bibitem[\protect\citeauthoryear{{Dav{\'e}}, {Angl{\'e}s-Alc{\'a}zar},
  {Narayanan}, {Li}, {Rafieferantsoa}  \& {Appleby}}{{Dav{\'e}}
  et~al.}{2019}]{Dave2019}
{Dav{\'e}} R.,  {Angl{\'e}s-Alc{\'a}zar} D.,  {Narayanan} D.,  {Li} Q.,
  {Rafieferantsoa} M.~H.,   {Appleby} S.,  2019, \mnras, \href
  {https://ui.adsabs.harvard.edu/abs/2019MNRAS.486.2827D} {486, 2827}

\bibitem[\protect\citeauthoryear{{Dolag}, {Borgani}, {Murante}  \&
  {Springel}}{{Dolag} et~al.}{2009}]{Dolag2009}
{Dolag} K.,  {Borgani} S.,  {Murante} G.,   {Springel} V.,  2009, \mnras, \href
  {http://adsabs.harvard.edu/abs/2009MNRAS.399..497D} {399, 497}

\bibitem[\protect\citeauthoryear{{Dubois} \& {Teyssier}}{{Dubois} \&
  {Teyssier}}{2008}]{Dubois2008}
{Dubois} Y.,  {Teyssier} R.,  2008, \aap, \href
  {https://ui.adsabs.harvard.edu/abs/2008A&A...482L..13D} {482, L13}

\bibitem[\protect\citeauthoryear{{Dubois} et~al.,}{{Dubois}
  et~al.}{2014}]{Dubois2014}
{Dubois} Y.,  et~al., 2014, \mnras, \href
  {https://ui.adsabs.harvard.edu/abs/2014MNRAS.444.1453D} {444, 1453}

\bibitem[\protect\citeauthoryear{{Dursi} \& {Pfrommer}}{{Dursi} \&
  {Pfrommer}}{2008}]{Dursi2008}
{Dursi} L.~J.,  {Pfrommer} C.,  2008, \apj, \href
  {https://ui.adsabs.harvard.edu/abs/2008ApJ...677..993D} {677, 993}

\bibitem[\protect\citeauthoryear{{Faucher-Gigu{\`e}re}, {Lidz}, {Zaldarriaga}
  \& {Hernquist}}{{Faucher-Gigu{\`e}re} et~al.}{2009}]{Faucher2009}
{Faucher-Gigu{\`e}re} C.-A.,  {Lidz} A.,  {Zaldarriaga} M.,   {Hernquist} L.,
  2009, \apj, \href {http://adsabs.harvard.edu/abs/2009ApJ...703.1416F} {703,
  1416}

\bibitem[\protect\citeauthoryear{{Ferri{\`e}re} \& {Terral}}{{Ferri{\`e}re} \&
  {Terral}}{2014}]{Ferriere2014}
{Ferri{\`e}re} K.,  {Terral} P.,  2014, \aap, \href
  {https://ui.adsabs.harvard.edu/abs/2014A&A...561A.100F} {561, A100}

\bibitem[\protect\citeauthoryear{{Ford}, {Dav{\'e}}, {Oppenheimer}, {Katz},
  {Kollmeier}, {Thompson}  \& {Weinberg}}{{Ford} et~al.}{2014}]{Ford2014}
{Ford} A.~B.,  {Dav{\'e}} R.,  {Oppenheimer} B.~D.,  {Katz} N.,  {Kollmeier}
  J.~A.,  {Thompson} R.,   {Weinberg} D.~H.,  2014, \mnras, \href
  {https://ui.adsabs.harvard.edu/abs/2014MNRAS.444.1260F} {444, 1260}

\bibitem[\protect\citeauthoryear{{Fumagalli}, {O'Meara}  \&
  {Prochaska}}{{Fumagalli} et~al.}{2016}]{Fumagalli2016}
{Fumagalli} M.,  {O'Meara} J.~M.,   {Prochaska} J.~X.,  2016, \mnras, \href
  {https://ui.adsabs.harvard.edu/abs/2016MNRAS.455.4100F} {455, 4100}

\bibitem[\protect\citeauthoryear{{Garaldi}, {Pakmor}  \& {Springel}}{{Garaldi}
  et~al.}{2020}]{Garaldi2020}
{Garaldi} E.,  {Pakmor} R.,   {Springel} V.,  2020, arXiv e-prints, \href
  {https://ui.adsabs.harvard.edu/abs/2020arXiv201009729G} {p. arXiv:2010.09729}

\bibitem[\protect\citeauthoryear{{Genel} et~al.,}{{Genel}
  et~al.}{2019}]{Genel2019}
{Genel} S.,  et~al., 2019, \apj, \href
  {https://ui.adsabs.harvard.edu/abs/2019ApJ...871...21G} {871, 21}

\bibitem[\protect\citeauthoryear{{Girichidis} et~al.,}{{Girichidis}
  et~al.}{2020}]{Girichidis2020}
{Girichidis} P.,  et~al., 2020, \ssr, \href
  {https://ui.adsabs.harvard.edu/abs/2020SSRv..216...68G} {216, 68}

\bibitem[\protect\citeauthoryear{{Grand} et~al.,}{{Grand}
  et~al.}{2017}]{Grand2017}
{Grand} R.~J.~J.,  et~al., 2017, \mnras, \href
  {http://adsabs.harvard.edu/abs/2017MNRAS.467..179G} {467, 179}

\bibitem[\protect\citeauthoryear{{Grand} et~al.,}{{Grand}
  et~al.}{2019}]{Grand2019}
{Grand} R. J.~J.,  et~al., 2019, \mnras, \href
  {https://ui.adsabs.harvard.edu/abs/2019MNRAS.490.4786G} {490, 4786}

\bibitem[\protect\citeauthoryear{{Gronke} \& {Oh}}{{Gronke} \&
  {Oh}}{2020}]{Gronke2020}
{Gronke} M.,  {Oh} S.~P.,  2020, \mnras, \href
  {https://ui.adsabs.harvard.edu/abs/2020MNRAS.492.1970G} {492, 1970}

\bibitem[\protect\citeauthoryear{{Gutcke}, {Stinson}, {Macci{\`o}}, {Wang}  \&
  {Dutton}}{{Gutcke} et~al.}{2017}]{Gutcke2017}
{Gutcke} T.~A.,  {Stinson} G.~S.,  {Macci{\`o}} A.~V.,  {Wang} L.,   {Dutton}
  A.~A.,  2017, \mnras, \href
  {https://ui.adsabs.harvard.edu/abs/2017MNRAS.464.2796G} {464, 2796}

\bibitem[\protect\citeauthoryear{{Han}}{{Han}}{2017}]{Han2017}
{Han} J.~L.,  2017, \araa, \href
  {https://ui.adsabs.harvard.edu/abs/2017ARA&A..55..111H} {55, 111}

\bibitem[\protect\citeauthoryear{{Haverkorn} \& {Heesen}}{{Haverkorn} \&
  {Heesen}}{2012}]{Haverkorn2012}
{Haverkorn} M.,  {Heesen} V.,  2012, \ssr, \href
  {https://ui.adsabs.harvard.edu/abs/2012SSRv..166..133H} {166, 133}

\bibitem[\protect\citeauthoryear{{Hennebelle} \& {Inutsuka}}{{Hennebelle} \&
  {Inutsuka}}{2019}]{Hennebelle2019}
{Hennebelle} P.,  {Inutsuka} S.-i.,  2019, Frontiers in Astronomy and Space
  Sciences, \href {https://ui.adsabs.harvard.edu/abs/2019FrASS...6....5H} {6,
  5}

\bibitem[\protect\citeauthoryear{{Hopkins} et~al.,}{{Hopkins}
  et~al.}{2018}]{Hopkins2018}
{Hopkins} P.~F.,  et~al., 2018, \mnras, \href
  {https://ui.adsabs.harvard.edu/abs/2018MNRAS.480..800H} {480, 800}

\bibitem[\protect\citeauthoryear{{Hopkins} et~al.,}{{Hopkins}
  et~al.}{2020}]{Hopkins2020}
{Hopkins} P.~F.,  et~al., 2020, \mnras, \href
  {https://ui.adsabs.harvard.edu/abs/2020MNRAS.492.3465H} {492, 3465}

\bibitem[\protect\citeauthoryear{{Hummels}, {Bryan}, {Smith}  \&
  {Turk}}{{Hummels} et~al.}{2013}]{Hummels2013}
{Hummels} C.~B.,  {Bryan} G.~L.,  {Smith} B.~D.,   {Turk} M.~J.,  2013, \mnras,
  \href {https://ui.adsabs.harvard.edu/abs/2013MNRAS.430.1548H} {430, 1548}

\bibitem[\protect\citeauthoryear{{Inoue}, {Hayashi}, {Shiota}, {Magara}  \&
  {Choe}}{{Inoue} et~al.}{2013}]{Inoue2013}
{Inoue} S.,  {Hayashi} K.,  {Shiota} D.,  {Magara} T.,   {Choe} G.~S.,  2013,
  \apj, \href {https://ui.adsabs.harvard.edu/abs/2013ApJ...770...79I} {770, 79}

\bibitem[\protect\citeauthoryear{{Jakobs} et~al.,}{{Jakobs}
  et~al.}{2018}]{Jakobs2018}
{Jakobs} A.,  et~al., 2018, \mnras, \href
  {https://ui.adsabs.harvard.edu/abs/2018MNRAS.480.3338J} {480, 3338}

\bibitem[\protect\citeauthoryear{{Ji}, {Oh}, {Ruszkowski}  \&
  {Markevitch}}{{Ji} et~al.}{2016}]{Ji2016}
{Ji} S.,  {Oh} S.~P.,  {Ruszkowski} M.,   {Markevitch} M.,  2016, \mnras, \href
  {https://ui.adsabs.harvard.edu/abs/2016MNRAS.463.3989J} {463, 3989}

\bibitem[\protect\citeauthoryear{{Ji}, {Oh}  \& {McCourt}}{{Ji}
  et~al.}{2018}]{Ji2018}
{Ji} S.,  {Oh} S.~P.,   {McCourt} M.,  2018, \mnras, \href
  {https://ui.adsabs.harvard.edu/abs/2018MNRAS.476..852J} {476, 852}

\bibitem[\protect\citeauthoryear{{Ji} et~al.,}{{Ji} et~al.}{2020}]{Ji2020}
{Ji} S.,  et~al., 2020, \mnras, \href
  {https://ui.adsabs.harvard.edu/abs/2020MNRAS.496.4221J} {496, 4221}

\bibitem[\protect\citeauthoryear{{Kacprzak}, {Churchill}  \&
  {Nielsen}}{{Kacprzak} et~al.}{2012}]{Kacprzak2012}
{Kacprzak} G.~G.,  {Churchill} C.~W.,   {Nielsen} N.~M.,  2012, \apjl, \href
  {https://ui.adsabs.harvard.edu/abs/2012ApJ...760L...7K} {760, L7}

\bibitem[\protect\citeauthoryear{{Kacprzak}, {Muzahid}, {Churchill}, {Nielsen}
  \& {Charlton}}{{Kacprzak} et~al.}{2015}]{Kacprzak2015}
{Kacprzak} G.~G.,  {Muzahid} S.,  {Churchill} C.~W.,  {Nielsen} N.~M.,
  {Charlton} J.~C.,  2015, \apj, \href
  {https://ui.adsabs.harvard.edu/abs/2015ApJ...815...22K} {815, 22}

\bibitem[\protect\citeauthoryear{{Kennicutt}}{{Kennicutt}}{1998}]{Kennicutt1998}
{Kennicutt} Robert~C. J.,  1998, \apj, \href
  {https://ui.adsabs.harvard.edu/abs/1998ApJ...498..541K} {498, 541}

\bibitem[\protect\citeauthoryear{{Lehner}, {Wotta}, {Howk},
  {O{\textquoteright}Meara}, {Oppenheimer}  \& {Cooksey}}{{Lehner}
  et~al.}{2019}]{Lehner2019}
{Lehner} N.,  {Wotta} C.~B.,  {Howk} J.~C.,  {O{\textquoteright}Meara} J.~M.,
  {Oppenheimer} B.~D.,   {Cooksey} K.~L.,  2019, \apj, \href
  {https://ui.adsabs.harvard.edu/abs/2019ApJ...887....5L} {887, 5}

\bibitem[\protect\citeauthoryear{{Li}, {Hopkins}, {Squire}  \& {Hummels}}{{Li}
  et~al.}{2020}]{Li2020}
{Li} Z.,  {Hopkins} P.~F.,  {Squire} J.,   {Hummels} C.,  2020, \mnras, \href
  {https://ui.adsabs.harvard.edu/abs/2020MNRAS.492.1841L} {492, 1841}

\bibitem[\protect\citeauthoryear{{Liang} \& {Remming}}{{Liang} \&
  {Remming}}{2020}]{Liang2020}
{Liang} C.~J.,  {Remming} I.,  2020, \mnras, \href
  {https://ui.adsabs.harvard.edu/abs/2020MNRAS.491.5056L} {491, 5056}

\bibitem[\protect\citeauthoryear{{Marinacci} \& {Vogelsberger}}{{Marinacci} \&
  {Vogelsberger}}{2016}]{Marinacci2016}
{Marinacci} F.,  {Vogelsberger} M.,  2016, \mnras, \href
  {https://ui.adsabs.harvard.edu/abs/2016MNRAS.456L..69M} {456, L69}

\bibitem[\protect\citeauthoryear{{Marinacci} et~al.,}{{Marinacci}
  et~al.}{2018}]{Marinacci2018}
{Marinacci} F.,  et~al., 2018, \mnras, \href
  {https://ui.adsabs.harvard.edu/abs/2018MNRAS.480.5113M} {480, 5113}

\bibitem[\protect\citeauthoryear{{Martin-Alvarez}, {Devriendt}, {Slyz}  \&
  {Teyssier}}{{Martin-Alvarez} et~al.}{2018}]{MartinAlvarez2018}
{Martin-Alvarez} S.,  {Devriendt} J.,  {Slyz} A.,   {Teyssier} R.,  2018,
  \mn@doi [\mnras] {10.1093/mnras/sty1623}, \href
  {https://ui.adsabs.harvard.edu/abs/2018MNRAS.479.3343M} {479, 3343}

\bibitem[\protect\citeauthoryear{{Martin-Alvarez}, {Slyz}, {Devriendt}  \&
  {G{\'o}mez-Guijarro}}{{Martin-Alvarez} et~al.}{2020}]{MartinAlvarez2020}
{Martin-Alvarez} S.,  {Slyz} A.,  {Devriendt} J.,   {G{\'o}mez-Guijarro} C.,
  2020, \mnras, \href {https://ui.adsabs.harvard.edu/abs/2020MNRAS.tmp.1559M}
  {}

\bibitem[\protect\citeauthoryear{{Martin}, {Ho}, {Kacprzak}  \&
  {Churchill}}{{Martin} et~al.}{2019}]{Martin2019}
{Martin} C.~L.,  {Ho} S.~H.,  {Kacprzak} G.~G.,   {Churchill} C.~W.,  2019,
  \apj, \href {https://ui.adsabs.harvard.edu/abs/2019ApJ...878...84M} {878, 84}

\bibitem[\protect\citeauthoryear{{McCourt}, {O'Leary}, {Madigan}  \&
  {Quataert}}{{McCourt} et~al.}{2015}]{McCourt2015}
{McCourt} M.,  {O'Leary} R.~M.,  {Madigan} A.-M.,   {Quataert} E.,  2015,
  \mnras, \href {https://ui.adsabs.harvard.edu/abs/2015MNRAS.449....2M} {449,
  2}

\bibitem[\protect\citeauthoryear{{Nelson} et~al.,}{{Nelson}
  et~al.}{2020}]{Nelson2020}
{Nelson} D.,  et~al., 2020, \mnras, \href
  {https://ui.adsabs.harvard.edu/abs/2020MNRAS.498.2391N} {498, 2391}

\bibitem[\protect\citeauthoryear{{Pakmor} \& {Springel}}{{Pakmor} \&
  {Springel}}{2013}]{Pakmor2013}
{Pakmor} R.,  {Springel} V.,  2013, \mnras, \href
  {https://ui.adsabs.harvard.edu/abs/2013MNRAS.432..176P} {432, 176}

\bibitem[\protect\citeauthoryear{{Pakmor}, {Marinacci}  \& {Springel}}{{Pakmor}
  et~al.}{2014}]{Pakmor2014}
{Pakmor} R.,  {Marinacci} F.,   {Springel} V.,  2014, \apjl, \href
  {https://ui.adsabs.harvard.edu/abs/2014ApJ...783L..20P} {783, L20}

\bibitem[\protect\citeauthoryear{{Pakmor}, {Springel}, {Bauer}, {Mocz},
  {Munoz}, {Ohlmann}, {Schaal}  \& {Zhu}}{{Pakmor} et~al.}{2016}]{Pakmor2016}
{Pakmor} R.,  {Springel} V.,  {Bauer} A.,  {Mocz} P.,  {Munoz} D.~J.,
  {Ohlmann} S.~T.,  {Schaal} K.,   {Zhu} C.,  2016, \mnras, \href
  {https://ui.adsabs.harvard.edu/abs/2016MNRAS.455.1134P} {455, 1134}

\bibitem[\protect\citeauthoryear{{Pakmor} et~al.,}{{Pakmor}
  et~al.}{2017}]{Pakmor2017}
{Pakmor} R.,  et~al., 2017, \mnras, \href
  {https://ui.adsabs.harvard.edu/abs/2017MNRAS.469.3185P} {469, 3185}

\bibitem[\protect\citeauthoryear{{Pakmor} et~al.,}{{Pakmor}
  et~al.}{2020}]{Pakmor2020}
{Pakmor} R.,  et~al., 2020, \mnras, \href
  {https://ui.adsabs.harvard.edu/abs/2020MNRAS.498.3125P} {498, 3125}

\bibitem[\protect\citeauthoryear{{P{\'e}roux}, {Nelson}, {van de Voort},
  {Pillepich}, {Marinacci}, {Vogelsberger}  \& {Hernquist}}{{P{\'e}roux}
  et~al.}{2020}]{Peroux2020}
{P{\'e}roux} C.,  {Nelson} D.,  {van de Voort} F.,  {Pillepich} A.,
  {Marinacci} F.,  {Vogelsberger} M.,   {Hernquist} L.,  2020, \mnras, \href
  {https://ui.adsabs.harvard.edu/abs/2020MNRAS.499.2462P} {499, 2462}

\bibitem[\protect\citeauthoryear{{Pillepich} et~al.,}{{Pillepich}
  et~al.}{2018}]{Pillepich2018}
{Pillepich} A.,  et~al., 2018, \mnras, \href
  {https://ui.adsabs.harvard.edu/abs/2018MNRAS.473.4077P} {473, 4077}

\bibitem[\protect\citeauthoryear{{Planck Collaboration} et~al.,}{{Planck
  Collaboration} et~al.}{2014}]{PlanckXVI2014}
{Planck Collaboration} et~al., 2014, \aap, \href
  {http://adsabs.harvard.edu/abs/2014A\%26A...571A..16P} {571, A16}

\bibitem[\protect\citeauthoryear{{Planck Collaboration} et~al.,}{{Planck
  Collaboration} et~al.}{2016}]{PlanckXXXV2016}
{Planck Collaboration} et~al., 2016, \aap, \href
  {https://ui.adsabs.harvard.edu/abs/2016A&A...586A.138P} {586, A138}

\bibitem[\protect\citeauthoryear{{Powell}, {Roe}, {Linde}, {Gombosi}  \& {De
  Zeeuw}}{{Powell} et~al.}{1999}]{Powell1999}
{Powell} K.~G.,  {Roe} P.~L.,  {Linde} T.~J.,  {Gombosi} T.~I.,   {De Zeeuw}
  D.~L.,  1999, Journal of Computational Physics, \href
  {https://ui.adsabs.harvard.edu/abs/1999JCoPh.154..284P} {154, 284}

\bibitem[\protect\citeauthoryear{{Prochaska} et~al.,}{{Prochaska}
  et~al.}{2017}]{Prochaska2017}
{Prochaska} J.~X.,  et~al., 2017, \apj, \href
  {http://adsabs.harvard.edu/abs/2017ApJ...837..169P} {837, 169}

\bibitem[\protect\citeauthoryear{{Putman}, {Peek}  \& {Joung}}{{Putman}
  et~al.}{2012}]{Putman2012}
{Putman} M.~E.,  {Peek} J.~E.~G.,   {Joung} M.~R.,  2012, \araa, \href
  {http://adsabs.harvard.edu/abs/2012ARA\%26A..50..491P} {50, 491}

\bibitem[\protect\citeauthoryear{{Rahmati}, {Pawlik}, {Rai{\v c}evi\`c}  \&
  {Schaye}}{{Rahmati} et~al.}{2013}]{Rahmati2013}
{Rahmati} A.,  {Pawlik} A.~H.,  {Rai{\v c}evi\`c} M.,   {Schaye} J.,  2013,
  \mnras, \href {http://adsabs.harvard.edu/abs/2013MNRAS.430.2427R} {430, 2427}

\bibitem[\protect\citeauthoryear{{Rieder} \& {Teyssier}}{{Rieder} \&
  {Teyssier}}{2017}]{Rieder2017}
{Rieder} M.,  {Teyssier} R.,  2017, \mnras, \href
  {https://ui.adsabs.harvard.edu/abs/2017MNRAS.472.4368R} {472, 4368}

\bibitem[\protect\citeauthoryear{{Sales}, {Navarro}, {Theuns}, {Schaye},
  {White}, {Frenk}, {Crain}  \& {Dalla Vecchia}}{{Sales}
  et~al.}{2012}]{Sales2012}
{Sales} L.~V.,  {Navarro} J.~F.,  {Theuns} T.,  {Schaye} J.,  {White} S. D.~M.,
   {Frenk} C.~S.,  {Crain} R.~A.,   {Dalla Vecchia} C.,  2012, \mnras, \href
  {https://ui.adsabs.harvard.edu/abs/2012MNRAS.423.1544S} {423, 1544}

\bibitem[\protect\citeauthoryear{{Schaye} et~al.,}{{Schaye}
  et~al.}{2015}]{Schaye2015}
{Schaye} J.,  et~al., 2015, \mnras, \href
  {http://adsabs.harvard.edu/abs/2015MNRAS.446..521S} {446, 521}

\bibitem[\protect\citeauthoryear{{Schroetter} et~al.,}{{Schroetter}
  et~al.}{2019}]{Schroetter2019}
{Schroetter} I.,  et~al., 2019, \mnras, \href
  {https://ui.adsabs.harvard.edu/abs/2019MNRAS.490.4368S} {490, 4368}

\bibitem[\protect\citeauthoryear{{Sharma}, {Parrish}  \& {Quataert}}{{Sharma}
  et~al.}{2010}]{Sharma2010}
{Sharma} P.,  {Parrish} I.~J.,   {Quataert} E.,  2010, \apj, \href
  {https://ui.adsabs.harvard.edu/abs/2010ApJ...720..652S} {720, 652}

\bibitem[\protect\citeauthoryear{{Simpson}, {Pakmor}, {Marinacci}, {Pfrommer},
  {Springel}, {Glover}, {Clark}  \& {Smith}}{{Simpson}
  et~al.}{2016}]{Simpson2016}
{Simpson} C.~M.,  {Pakmor} R.,  {Marinacci} F.,  {Pfrommer} C.,  {Springel} V.,
   {Glover} S. C.~O.,  {Clark} P.~C.,   {Smith} R.~J.,  2016, \apjl, \href
  {https://ui.adsabs.harvard.edu/abs/2016ApJ...827L..29S} {827, L29}

\bibitem[\protect\citeauthoryear{{Sparre}, {Pfrommer}  \& {Ehlert}}{{Sparre}
  et~al.}{2020}]{Sparre2020}
{Sparre} M.,  {Pfrommer} C.,   {Ehlert} K.,  2020, \mnras, \href
  {https://ui.adsabs.harvard.edu/abs/2020MNRAS.499.4261S} {499, 4261}

\bibitem[\protect\citeauthoryear{{Springel}}{{Springel}}{2010}]{Springel2010}
{Springel} V.,  2010, \mnras, \href
  {http://adsabs.harvard.edu/abs/2010MNRAS.401..791S} {401, 791}

\bibitem[\protect\citeauthoryear{{Springel} \& {Hernquist}}{{Springel} \&
  {Hernquist}}{2003}]{Springel2003}
{Springel} V.,  {Hernquist} L.,  2003, \mnras, \href
  {https://ui.adsabs.harvard.edu/abs/2003MNRAS.339..289S} {339, 289}

\bibitem[\protect\citeauthoryear{{Springel}, {White}, {Tormen}  \&
  {Kauffmann}}{{Springel} et~al.}{2001}]{Springel2001}
{Springel} V.,  {White} S.~D.~M.,  {Tormen} G.,   {Kauffmann} G.,  2001,
  \mnras, \href {http://adsabs.harvard.edu/abs/2001MNRAS.328..726S} {328, 726}

\bibitem[\protect\citeauthoryear{{Su}, {Hopkins}, {Hayward},
  {Faucher-Gigu{\`e}re}, {Kere{\v{s}}}, {Ma}  \& {Robles}}{{Su}
  et~al.}{2017}]{Su2017}
{Su} K.-Y.,  {Hopkins} P.~F.,  {Hayward} C.~C.,  {Faucher-Gigu{\`e}re} C.-A.,
  {Kere{\v{s}}} D.,  {Ma} X.,   {Robles} V.~H.,  2017, \mnras, \href
  {https://ui.adsabs.harvard.edu/abs/2017MNRAS.471..144S} {471, 144}

\bibitem[\protect\citeauthoryear{{Suresh}, {Bird}, {Vogelsberger}, {Genel},
  {Torrey}, {Sijacki}, {Springel}  \& {Hernquist}}{{Suresh}
  et~al.}{2015}]{Suresh2015}
{Suresh} J.,  {Bird} S.,  {Vogelsberger} M.,  {Genel} S.,  {Torrey} P.,
  {Sijacki} D.,  {Springel} V.,   {Hernquist} L.,  2015, \mnras, \href
  {https://ui.adsabs.harvard.edu/abs/2015MNRAS.448..895S} {448, 895}

\bibitem[\protect\citeauthoryear{{Tchekhovskoy}}{{Tchekhovskoy}}{2015}]{Tchekhovskoy2015}
{Tchekhovskoy} A.,  2015, in {Contopoulos} I.,  {Gabuzda} D.,   {Kylafis} N.,
  eds,  Astrophysics and Space Science Library Vol. 414, The Formation and
  Disruption of Black Hole Jets. Springer International Publishing, Cham,
  Switzerland, p.~45

\bibitem[\protect\citeauthoryear{{Tremmel}, {Karcher}, {Governato},
  {Volonteri}, {Quinn}, {Pontzen}, {Anderson}  \& {Bellovary}}{{Tremmel}
  et~al.}{2017}]{Tremmel2017}
{Tremmel} M.,  {Karcher} M.,  {Governato} F.,  {Volonteri} M.,  {Quinn} T.~R.,
  {Pontzen} A.,  {Anderson} L.,   {Bellovary} J.,  2017, \mnras, \href
  {https://ui.adsabs.harvard.edu/abs/2017MNRAS.470.1121T} {470, 1121}

\bibitem[\protect\citeauthoryear{{Tumlinson}, {Peeples}  \& {Werk}}{{Tumlinson}
  et~al.}{2017}]{Tumlinson2017}
{Tumlinson} J.,  {Peeples} M.~S.,   {Werk} J.~K.,  2017, \araa, \href
  {http://adsabs.harvard.edu/abs/2017ARA\%26A..55..389T} {55, 389}

\bibitem[\protect\citeauthoryear{{van de Voort} \& {Schaye}}{{van de Voort} \&
  {Schaye}}{2012}]{Voort2012}
{van de Voort} F.,  {Schaye} J.,  2012, \mnras, \href
  {https://ui.adsabs.harvard.edu/abs/2012MNRAS.423.2991V} {423, 2991}

\bibitem[\protect\citeauthoryear{{van de Voort}, {Springel}, {Mandelker}, {van
  den Bosch}  \& {Pakmor}}{{van de Voort} et~al.}{2019}]{Voort2019}
{van de Voort} F.,  {Springel} V.,  {Mandelker} N.,  {van den Bosch} F.~C.,
  {Pakmor} R.,  2019, \mnras, \href
  {https://ui.adsabs.harvard.edu/abs/2019MNRAS.482L..85V} {482, L85}

\bibitem[\protect\citeauthoryear{{van de Voort}, {Pakmor}, {Grand}, {Springel},
  {G{\'o}mez}  \& {Marinacci}}{{van de Voort} et~al.}{2020}]{Voort2020}
{van de Voort} F.,  {Pakmor} R.,  {Grand} R. J.~J.,  {Springel} V.,
  {G{\'o}mez} F.~A.,   {Marinacci} F.,  2020, \mnras, \href
  {https://ui.adsabs.harvard.edu/abs/2020MNRAS.494.4867V} {494, 4867}

\bibitem[\protect\citeauthoryear{{Vogelsberger}, {Genel}, {Sijacki}, {Torrey},
  {Springel}  \& {Hernquist}}{{Vogelsberger} et~al.}{2013}]{Vogelsberger2013}
{Vogelsberger} M.,  {Genel} S.,  {Sijacki} D.,  {Torrey} P.,  {Springel} V.,
  {Hernquist} L.,  2013, \mnras, \href
  {https://ui.adsabs.harvard.edu/abs/2013MNRAS.436.3031V} {436, 3031}

\bibitem[\protect\citeauthoryear{{Weinberger} et~al.,}{{Weinberger}
  et~al.}{2017}]{Weinberger2017}
{Weinberger} R.,  et~al., 2017, \mnras, \href
  {https://ui.adsabs.harvard.edu/abs/2017MNRAS.465.3291W} {465, 3291}

\bibitem[\protect\citeauthoryear{{Weinberger}, {Springel}  \&
  {Pakmor}}{{Weinberger} et~al.}{2020}]{Weinberger2020}
{Weinberger} R.,  {Springel} V.,   {Pakmor} R.,  2020, \apjs, \href
  {https://ui.adsabs.harvard.edu/abs/2020ApJS..248...32W} {248, 32}

\bibitem[\protect\citeauthoryear{{Werk} et~al.,}{{Werk}
  et~al.}{2014}]{Werk2014}
{Werk} J.~K.,  et~al., 2014, \apj, \href
  {http://adsabs.harvard.edu/abs/2014ApJ...792....8W} {792, 8}

\bibitem[\protect\citeauthoryear{{Whittingham}, {Sparre}, {Pfrommer}  \&
  {Pakmor}}{{Whittingham} et~al.}{2020}]{Whittingham2020}
{Whittingham} J.,  {Sparre} M.,  {Pfrommer} C.,   {Pakmor} R.,  2020, arXiv
  e-prints, \href {https://ui.adsabs.harvard.edu/abs/2020arXiv201113947W} {p.
  arXiv:2011.13947}

\bibitem[\protect\citeauthoryear{{Wiener}, {Zweibel}  \& {Ruszkowski}}{{Wiener}
  et~al.}{2019}]{Wiener2019}
{Wiener} J.,  {Zweibel} E.~G.,   {Ruszkowski} M.,  2019, \mnras, \href
  {https://ui.adsabs.harvard.edu/abs/2019MNRAS.489..205W} {489, 205}

\bibitem[\protect\citeauthoryear{{Zahedy}, {Chen}, {Johnson}, {Pierce},
  {Rauch}, {Huang}, {Weiner}  \& {Gauthier}}{{Zahedy}
  et~al.}{2019}]{Zahedy2019}
{Zahedy} F.~S.,  {Chen} H.-W.,  {Johnson} S.~D.,  {Pierce} R.~M.,  {Rauch} M.,
  {Huang} Y.-H.,  {Weiner} B.~J.,   {Gauthier} J.-R.,  2019, \mnras, \href
  {https://ui.adsabs.harvard.edu/abs/2019MNRAS.484.2257Z} {484, 2257}

\makeatother
\end{thebibliography}

\bsp

\appendix

\normalsize

\section{Resolution tests} \label{sec:test}

In order to test for resolution effects, we have repeated our six fiducial simulations without CGM refinement (i.e.\ the standard simulation method with mass refinement only). The difference in CGM resolution depends on the density of the gas. Gas with densities $n_\mathrm{H}>0.0017$ have the same resolution as before. At lower densities the mass resolution remains the same, which means that the spatial resolution worsens substantially without CGM refinement. Because the density generally decreases with galactocentric radius, the ISM and innermost CGM are treated in the same way in both refinement methods, but the resolution of the CGM at $R_\mathrm{GC}>20-30$~kpc is, on average, worse with mass refinement only. The resolution difference increases towards larger radii. At $R_\mathrm{GC}=200$~kpc, the mass resolution of our spatially refined simulation is 2 orders of magnitude better than standard simulations with mass refinement only. In this Appendix we reproduce some of our main results based on the mass refinement only simulations to show that the impact of magnetic fields described above are robust to changes in CGM resolution. 

\begin{table*}
\begin{center}
\caption{\label{tab:res} \small Properties of the galaxy and halo at $z=0$, unless otherwise stated, in our simulations with standard resolution: galaxy formation model, inclusion of $B$ field, total stellar mass within 30~kpc from the centre ($M_\mathrm{star}$), mass of the central black hole ($M_\mathrm{BH}$), total ISM mass within 30~kpc from the centre ($M_\mathrm{ISM}$), total CGM mass (i.e.\ all non-star-forming gas within $R_\mathrm{vir}$; $M_\mathrm{CGM}$), the total amount of H~\textsc{i} in the CGM ($M_\mathrm{HI}$), star formation rate averaged over $z=0.3-0$ (SFR), and kinetic disc-to-total ratio based on $\kappa_\mathrm{rot}$ (D/T; see Equation~\ref{eqn:kappa}). The effect of magnetic fields on these properties at standard resolution is very similar to those in our simulations with CGM refinement (see Table~\ref{tab:prop}). We therefore consider these results robust to changes in CGM resolution.}
\vspace{-4mm}
\begin{tabular}[t]{lrrrrrrcr}
\hline \\[-3mm]
simulation & $B$ & log$_{10}$ $M_\mathrm{star}$ & log$_{10}$ $M_\mathrm{BH}$ & log$_{10}$ $M_\mathrm{ISM}$ & log$_{10}$ $M_\mathrm{CGM}$ & log$_{10}$ $M_\mathrm{HI}$ & $\langle$SFR$\rangle$                                     & D/T \\
name         & field         & [M$_{\astrosun}$] & [M$_{\astrosun}$]  &  [M$_{\astrosun}$]  & [M$_{\astrosun}$]   & [M$_{\astrosun}$]    & (M$_{\astrosun}$~yr$^{-1}$)  & ($\kappa_\mathrm{rot}$)  \\
\hline \\[-4mm]                                                                                                                                       
\color{red} Auriga noAGN    & \color{red} no    & \color{red} ${10.82}$  & \color{red} --                 & \color{red} ${9.82}$   & \color{red} ${10.48}$  & \color{red} ${9.81}$   & \color{red} 4.9   & \color{red} 0.62  \\
\color{blue} Auriga noAGN  & \color{blue} yes & \color{blue} ${10.91}$ & \color{blue} --               & \color{blue} ${9.82}$ & \color{blue} ${10.79}$ & \color{blue} ${10.49}$  & \color{blue} 6.0 & \color{blue} 0.70  \\[1mm]
\color{red} Auriga                & \color{red} no    & \color{red} ${10.70}$  & \color{red} ${7.39}$   & \color{red} ${9.71}$   & \color{red} ${10.64}$   & \color{red} ${9.81}$  & \color{red} 3.6   & \color{red} 0.62  \\
\color{blue} Auriga              & \color{blue} yes & \color{blue} ${10.71}$ & \color{blue} ${7.51}$ & \color{blue} ${9.45}$ & \color{blue} ${10.75}$ & \color{blue} ${10.34}$  & \color{blue} 3.7 & \color{blue} 0.71  \\[1mm]
\color{red} TNG                   & \color{red} no    & \color{red} ${10.61}$   & \color{red} ${7.46}$   & \color{red} ${9.68}$   & \color{red} ${10.61}$   & \color{red} ${10.07}$  & \color{red} 2.6   & \color{red} 0.73  \\
\color{blue} TNG                 & \color{blue} yes & \color{blue} ${10.59}$  & \color{blue} ${7.72}$ & \color{blue} ${9.62}$ & \color{blue} ${10.74}$ & \color{blue} ${10.17}$  & \color{blue} 2.2 & \color{blue} 0.76  \\[-1mm]
\hline
\end{tabular}
\end{center}
\end{table*}   

Table~\ref{tab:res} is the equivalent of Table~\ref{tab:prop} for the standard resolution simulations rather than those with 1~kpc spatial refinement. Owing to the chaotic nature of galaxy formation, we do not expect results to be identical. Stellar masses are within 0.04~dex of one another for the simulations with different refinement methods and black hole masses within 0.07~dex, consistent with expectations from the butterfly effect \citep{Genel2019}. The differences and similarities between properties with and without magnetic fields described in Section~\ref{sec:results} are reproduced by the simulations without CGM refinement. Specifically, the presence of magnetic fields leads to higher black hole masses, CGM masses, and H\,\textsc{i} masses, but no clear trend in ISM masses and SFRs, and an increase in disc-to-total ratio. 

\begin{figure}
\center
\includegraphics[scale=0.59]{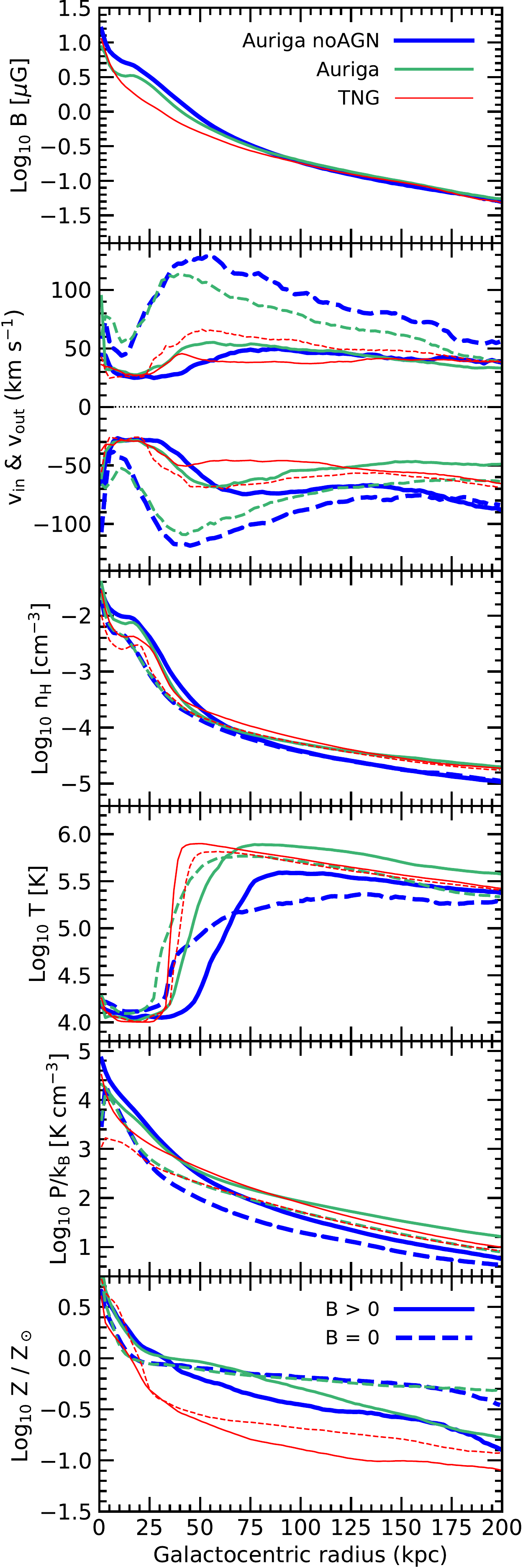}
\caption {\label{fig:res} Profiles of CGM properties as in Figure~\ref{fig:prof}, but for simulations with standard resolution (i.e.\ mass refinement only).}
\end{figure} 

The properties of the CGM in our simulations with mass refinement only are shown in Figure~\ref{fig:res}, which can be compared to those in Figure~\ref{fig:prof}. The minor differences between the two figures are likely primarily due to the stochastic nature of galaxy formation, given that the curves are based on a single halo (though averaged over 22 simulation outputs from $z=0.3$ to $z=0$). The important thing to note is that the changes in CGM properties caused by the presence of magnetic fiels are very similar even at this lower spatial resolution in the CGM. As seen previously for our main simulations: with $B$ fields both gas inflows and outflows have lower velocities, the density is higher in the inner CGM (at $R_\mathrm{GC}\lesssim50$~kpc), the temperature is lower at $R_\mathrm{GC}\gtrsim50$~kpc, the (thermal + magnetic) pressure is higher, and the metallcity is lower at $R_\mathrm{GC}\gtrsim50$~kpc. We therefore consider our results converged with respect to CGM resolution. 

The magnitude of the changes due to magnetic fields varies with galaxy formation models and is generally largest for `Auriga noAGN' and smaller for the two models with AGN feedback, as also found for our CGM refined simulations. This could be due to additional feedback mitigating some of the effects that magnetic fields have on the circumgalactic gas flows. We therefore again find that the results are qualitatively unchanged for the various galaxy formation models. 

There is one noticeable change with resolution when comparing the bottom panels of Figures~\ref{fig:res} and~\ref{fig:prof}. The difference in metallicity between simulations with and without magnetic fields is somewhat larger in the higher resolution simulations. This could be because numerical mixing is reduced in our spatially refined simulations. It is therefore possible that the influence of magnetic fields on the CGM metallicity will continue to increase with improved resolution.

\section{Halo-to-halo variation} \label{sec:haloes}

\begin{table*}
\begin{center}
\caption{\label{tab:haloes} \small Properties of the galaxy and halo at $z=0$, unless otherwise stated, in our simulations with standard resolution: halo identifier from the Auriga suite \citep{Grand2017, Grand2019}, inclusion of $B$ field, total mass within $R_\mathrm{vir}$ ($M_\mathrm{halo}$), total stellar mass within 30~kpc from the centre ($M_\mathrm{star}$), total ISM mass within 30~kpc from the centre ($M_\mathrm{ISM}$), total CGM mass (i.e.\ all non-star-forming gas within $R_\mathrm{vir}$; $M_\mathrm{CGM}$), the total amount of H~\textsc{i} in the CGM ($M_\mathrm{HI}$), star formation rate averaged over $z=0.3-0$ (SFR), and kinetic disc-to-total ratio based on $\kappa_\mathrm{rot}$ (D/T; see Equation~\ref{eqn:kappa}). The same qualitative change in properties caused by the inclusion of magnetic fields is seen for all three haloes. Although this is still a limited sample of haloes, it shows that the effects of magnetic fields do not depend purely on the chosen initial conditions. We therefore conclude that our findings are likely robust for Milky Way-mass haloes.}
\vspace{-4mm}
\begin{tabular}[t]{lrrrrrrcr}
\hline \\[-3mm]
simulation & $B$ & log$_{10}$ $M_\mathrm{halo}$ & log$_{10}$ $M_\mathrm{star}$ & log$_{10}$ $M_\mathrm{ISM}$ & log$_{10}$ $M_\mathrm{CGM}$ & log$_{10}$ $M_\mathrm{HI}$ & $\langle$SFR$\rangle$                                     & D/T \\
name         & field         & [M$_{\astrosun}$] & [M$_{\astrosun}$]  &  [M$_{\astrosun}$]  & [M$_{\astrosun}$]   & [M$_{\astrosun}$]    & (M$_{\astrosun}$~yr$^{-1}$)  & ($\kappa_\mathrm{rot}$)  \\
\hline \\[-4mm]                                                                                                                                       
\color{red} halo 6    & \color{red} no    & \color{red} ${12.01}$  & \color{red} ${10.82}$  & \color{red} ${9.82}$   & \color{red} ${10.48}$  & \color{red} ${9.81}$   & \color{red} 4.9   & \color{red} 0.62  \\
\color{blue} halo 6  & \color{blue} yes & \color{blue} ${12.04}$ & \color{blue} ${10.91}$ & \color{blue} ${9.82}$ & \color{blue} ${10.79}$ & \color{blue} ${10.49}$  & \color{blue} 6.0 & \color{blue} 0.70  \\[1mm]
\color{red} halo 12  & \color{red} no    & \color{red} ${12.04}$  & \color{red} ${10.94}$   & \color{red} ${10.13}$   & \color{red} ${10.54}$   & \color{red} ${9.76}$  & \color{red} 12.2   & \color{red} 0.60  \\
\color{blue} halo 12 & \color{blue} yes & \color{blue} ${12.07}$ & \color{blue} ${11.01}$ & \color{blue} ${10.47}$ & \color{blue} ${10.77}$ & \color{blue} ${10.38}$  & \color{blue} 19.2 & \color{blue} 0.67  \\[1mm]
\color{red} halo L8   & \color{red} no    & \color{red} ${11.92}$   & \color{red} ${10.76}$   & \color{red} ${10.21}$ & \color{red} ${10.59}$ & \color{red} ${10.02}$ & \color{red} 7.9  & \color{red} 0.70  \\
\color{blue} halo L8 & \color{blue} yes & \color{blue} ${11.96}$  & \color{blue} ${10.91}$ & \color{blue} ${10.35}$ & \color{blue} ${10.74}$ & \color{blue} ${10.42}$  & \color{blue} 11.9 & \color{blue} 0.74  \\[-1mm]
\hline
\end{tabular}
\end{center}
\end{table*}   

In order to get an idea of the variation between different haloes, we ran two additional sets of simulations at standard resolution (i.e.\ mass refinement only) as in Appendix~\ref{sec:test}. Halo~6 is our fiducial halo on which our main results are based. Halo~12 is also part of the orignal Auriga suite and its properties are described in \citet{Grand2017}, along with those of halo~6. Halo~L8 is part of an extension of the original suite to slightly lower halo masses and described in \citet{Grand2019} and \citet{Voort2020}. We ran each halo twice, once with and once without magnetic fields, both using model `Auriga noAGN'. Some of the resulting galaxy and halo properties are listed in Table~\ref{tab:haloes}. 

For all three haloes, the differences between the simulations with or without magnetic fields, which we identified previously based solely on halo~6, are reproduced. With $B>0$, the total CGM mass, the H~\textsc{i} mass in the CGM and the disc-to-total ratio of the central galaxy are all higher than when $B=0$. We therefore conclude that these results are likely robust for Milky Way-mass haloes, although a larger sample would be useful to test this further. The new haloes~12 and~L8 also show an increase in stellar mass, ISM mass, and SFR, but we previously saw several counterexamples in our other simulations (see Tables~\ref{tab:prop} and~\ref{tab:res}), so we do not consider these changes to be robust. 

\begin{figure}
\center
\includegraphics[scale=0.59]{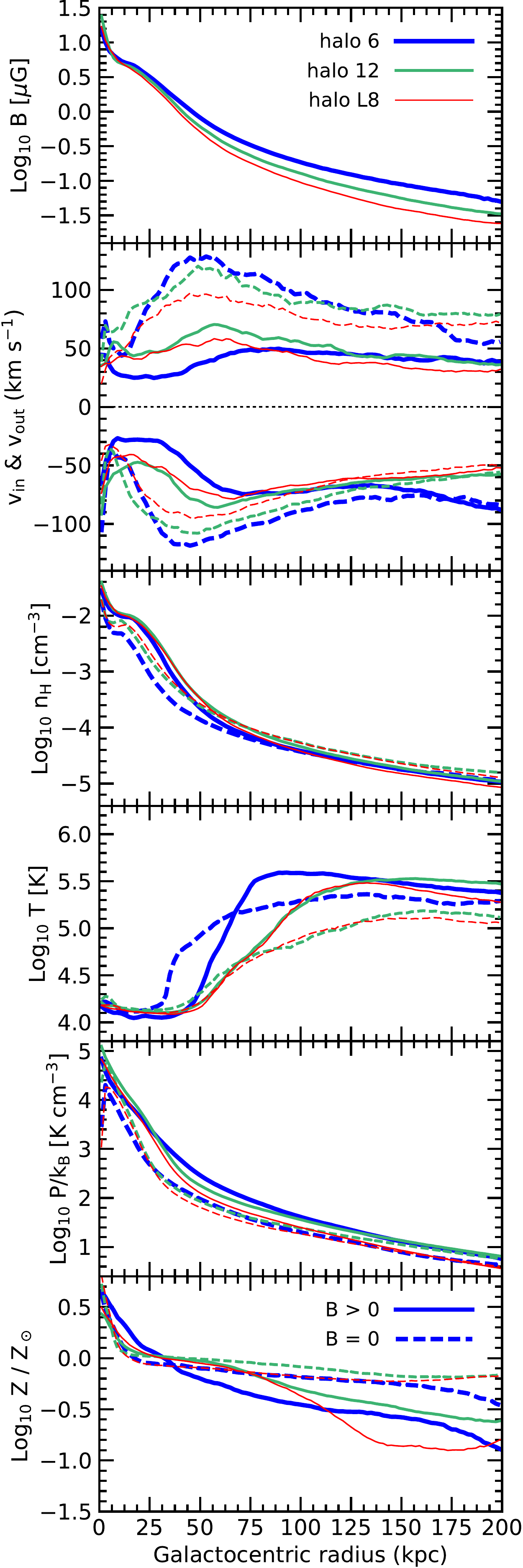}
\caption {\label{fig:haloes} Profiles of CGM properties as in Figure~\ref{fig:res} for 3 different haloes: halo 6 (thick, blue curves), halo 12 (green curves), and halo L8 (thin, red curves). All haloes were simulated with the `Auriga noAGN' model with $B>0$ (solid curves) and $B=0$ (dashed curves).}
\end{figure} 

Figure~\ref{fig:haloes} quantifies the CGM properties of the three different Milky Way-mass haloes as a function of galactocentric radius, as before in Figures~\ref{fig:prof} and~\ref{fig:res}. There are noticeable differences between the three haloes, but here we focus on their similarities. All three haloes show the expected decrease in volume-weighted root-mean-square magnetic field strength with galactocentric radius. With magnetic fields included, the inflow and outflow velocities are lower, on average. The central 50~kpc of the CGM is denser when $B>0$ for all three haloes. In the outer CGM, the median temperature of the gas is somewhat higher when magnetic fields are included. The total (thermal + magnetic) pressure is higher in the entire halo when $B>0$. For each of the three haloes, the simulation without magnetic fields shows an almost flat metallicity profile out to 200~kpc from the central galaxy. With magnetic fields, on the other hand, there is a clear decrease, resulting in much lower median metallicities in the outer CGM for each halo. This shows that our conclusions from the main body of this work, based on high-resolution simulations of halo~6, are qualitatively reproduced and do not depend purely on the initial conditions chosen. We therefore conclude that our results appear to be robust for Milky Way-mass haloes, based on this small sample.

\section{Probability density functions} \label{sec:pdf}

\begin{figure*}
\center
\includegraphics[scale=0.65]{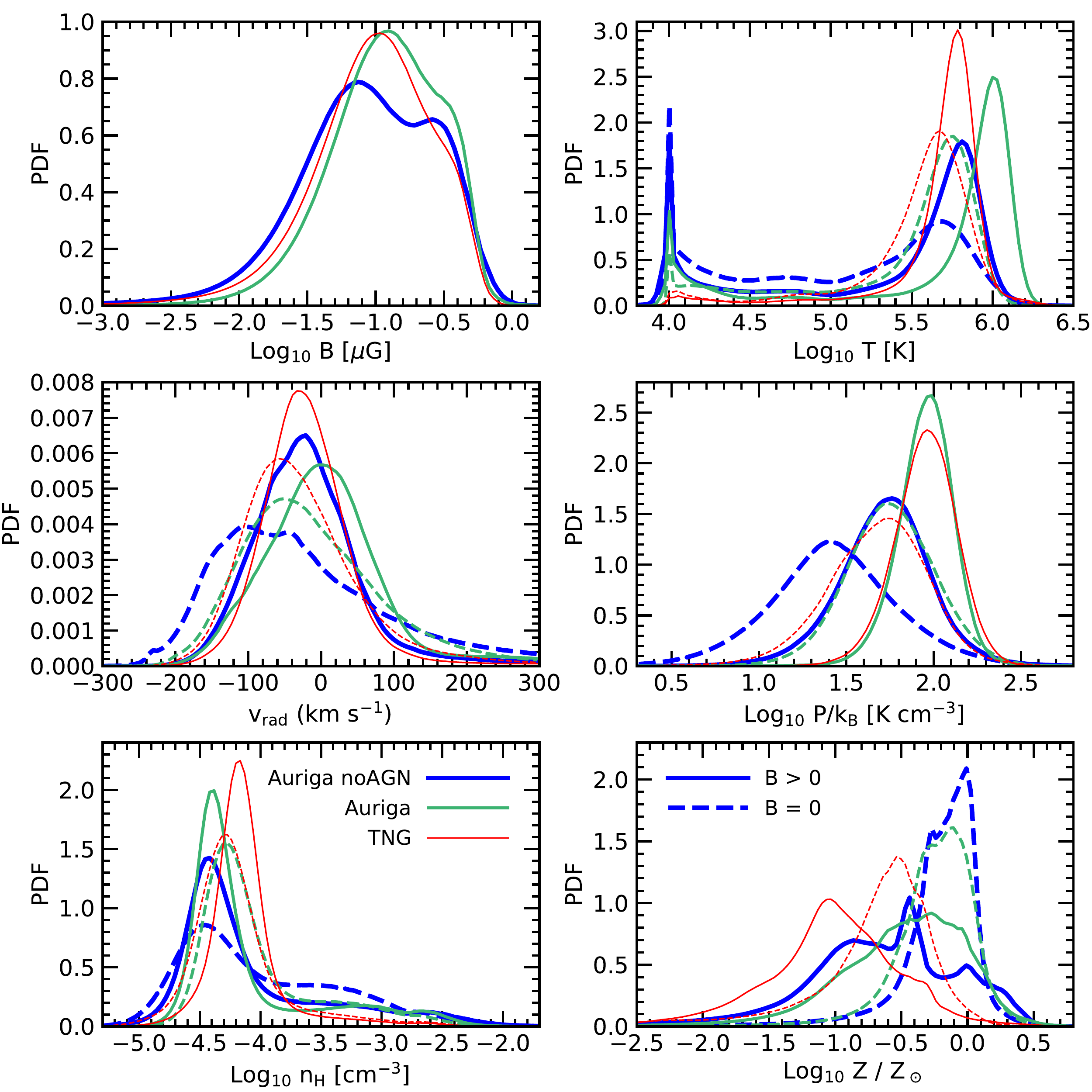}
\caption {\label{fig:pdf} Probability density functions for properties of the CGM at $90<R_\mathrm{GC}<110$~kpc for our high-resolution simulations with (solid curves) and without (dashed curves) magnetic fields for the three different galaxy formation models (as in Figure~\ref{fig:prof}). The left-hand column shows the magnetic field strength (top), the radial velocity (middle), and the hydrogen number density (bottom) and the right-hand column shows the temperature (top), pressure (middle), and metallicity (bottom). The distributions of radial velocities, densities, and pressures are narrower in the presence of magnetic fields, whereas the metallicity distribution is broader. The temperature distribution of the hot gas is also narrower and the importance of intermediate-temperature gas is reduced when $B>0$. Including magnetic fields results in a smoother CGM (less variation in density, temperature, and pressure) with slower outflows and inflows (reduced scatter in the radial velocity distribution) and less efficient mixing (enhanced scatter in the metallicity distribution at fixed galactocentric radius).}
\end{figure*} 

The distribution of CGM properties in a 20~kpc thick shell centred on the main galaxy, at galactocentric distances between 90 and 110~kpc, are shown as probability density functions (PDFs) in Figure~\ref{fig:pdf} for all six simulations with CGM refinement. Those with (without) magnetic fields are shown as solid (dashed) curves for models `Auriga noAGN' (thick, blue curves), `Auriga' (green curves), and `TNG' (thin, red curves). We show properties in a relatively thin shell to emphasize the scatter at `fixed' galactocentric distance. Gas associated with satellites has been excluded. To reduce the impact of stochastic processes, we show properties averaged over all 22 simulation outputs between $z=0.3$ and $z=0$. Below we describe the (qualitative) commonalities between all pairs of simulations (i.e.\ with and without magnetic fields) and have verified that the same behaviour is seen for our standard resolution simulations (shown in Appendix~\ref{sec:test}) and for the two simulations with different initial conditions (shown in Appendix~\ref{sec:haloes}). We therefore believe the described differences are robust to moderate changes in galaxy formation model, resolution, and initial conditions. 

The distribution of the magnetic field strength (top, left-hand panel) is relatively broad, spanning more than an order of magnitude. It also shows a tail towards low values. A small hint of a bimodality is visible, most prominently for simulation `Auriga noAGN'. The higher values correspond to the gas located along the angular momentum vector of the galaxy, where the gas has been enriched by outflows emanating from the central galaxy. The lower values correspond to the gas located away from the cones of outflowing gas, as can also be seen in Figure~\ref{fig:imgBbeta}.

The radial velocity distribution (middle, left-hand panel) is more peaked when $B>0$ and much broader when $B=0$. High inflow and outflow velocities are more likely to be found in simulations without magnetic fields. The peak of the distribution shifts to higher (less negative) velocities in all three galaxy formation models when magnetic fields are included.

The density distribution (bottom, left-hand panel) peaks at low densities and exhibits a prominent tail towards higher densities for all of our simulations. When $B>0$ the distribution has a stronger peak at low values. The three galaxy formation models show some differences, but all show a decrease in the importance of the high-density tail when magnetic fields are included. 

The distribution of temperatures (top, right-hand panel) is bimodal, showing a peak at the minimum CGM temperature (there is no radiative cooling below $10^4$~K in our simulations) and around the virial temperature of the halo ($T_\mathrm{vir}\approx10^6$~K). The peak at high temperature is stronger and shifted to slightly higher temperature when magnetic fields are included in the simulations, whilst less of the gas is at intermediate temperatures ($10^{4.2}\lesssim T \lesssim10^{5.5}$~K). Even though it is tempting to conclude from Figure~\ref{fig:imgBnoB} that magnetic fields increase the amount of cool to intermediate-temperature gas, because of the low-temperature filaments that are clearly visible in the image, the opposite is true. The cool filaments have a reasonably low density and the amount of cool gas actually decreases when $B>0$. This is likely connected to the decrease in the amount of relatively dense gas as seen in the bottom, left-hand panel. 

The pressure distribution (middle, right-hand panel) confirms our conclusion based on the smoother appearance of the pressure in Figure~\ref{fig:imgBnoB}. When magnetic fields are included, the distribution of pressures becomes more strongly peaked, thus showing reduced scatter at `fixed' distance from the central galaxy. This reduction of pressure variations proves that the CGM is indeed smoother when $B>0$. 

The metallicity distribution (bottom, right-hand panel) is strongly affected by the inclusion of magnetic fields. When $B=0$ the metallicity is strongly peaked at relatively high metallicities, whereas for $B>0$ the distribution is much wider and has a much larger contribution of low-metallicity gas. This is consistent with what we found in Figure~\ref{fig:angle}, i.e. the metals are more evenly distributed throughout the CGM without magnetic fields and less well-mixed when magnetic fields are included. 

All of these properties taken together indicate that including magnetic fields results in a smoother CGM (less variation in density, temperature, and pressure at fixed galactocentric radius), where outflows and inflows are slower (outliers in the radial velocity distribution are strongly reduced) and the gas mixes less efficiently (enhanced scatter in the metallicity distribution at fixed galactocentric radius).

\label{lastpage}

\end{document}